\documentclass[
 reprint,
 superscriptaddress,
 amsmath,amssymb,
 aps,
prl,
floatfix,
floatfix,]{revtex4-2}

\usepackage{comment}
\usepackage{graphicx} % Required for inserting images
\usepackage{color}
\usepackage{graphicx}% Include figure files
\usepackage{dcolumn}% Align table columns on decimal point
\usepackage{bm}% bold math
\usepackage{bbm} % for \mathbbm{1} identity operator
\usepackage[hidelinks]{hyperref}% add hypertext capabilities
\usepackage{siunitx}
\usepackage{mathtools} % for \DeclarePairedDelimiterX
\usepackage[capitalise]{cleveref} % for clever referencing use \cref{fig:...}(a) to display "Fig. 1(b)" instead of Fig.\,\ref{fig:...}(a)
\crefname{section}{Sec.}{Secs.} % ensure section refs have non-empty text, avoids empty hyperref link
\usepackage{placeins} % provides \FloatBarrier to force placement of pending floats

\usepackage{bibunits}
\defaultbibliography{Biblio}         % This sets the default .bib file
\defaultbibliographystyle{apsrev4-2}  % This sets the bibliography style

\newcommand{\e}{\textrm{e}}
\newcommand{\mi}{\mathrm{i}}

\newcommand{\dd}{\textrm{d}}

\DeclarePairedDelimiterX\abs[1]{\lvert}{\rvert}{#1}
\DeclarePairedDelimiterX\ket[1]{\lvert}{\rangle}{#1}
\DeclarePairedDelimiterX\bra[1]{\langle}{\rvert}{#1}
\DeclarePairedDelimiterX\braket[2]{\langle}{\rangle}{#1\,\vert\,#2}

\DeclarePairedDelimiterX\expval[2]{\langle}{\rangle}{#1\,\vert\,#2\,\vert\,#1}
\DeclarePairedDelimiterX\matel[3]{\langle}{\rangle}{#1\,\vert\,#2\,\vert\,#3}
\DeclarePairedDelimiterX\ep[1]{\langle}{\rangle}{#1}

\DeclarePairedDelimiterX\ketbra[2]{\lvert}{\rvert}{#1\rangle\langle#2}
\DeclarePairedDelimiterX\projector[1]{\lvert}{\rvert}{#1\rangle\langle#1}
\newcommand{\del}{\partial} % partial derivative
\begin{document}

\title{Revealing Pseudo-Fermionization and Chiral Binding of One-Dimensional Anyons using Adiabatic State Preparation}

\author{Brice Bakkali-Hassani}
\altaffiliation[Current address: ]{Laboratoire Kastler Brossel, Coll\`ege de France, CNRS, ENS-Universit\'e PSL, Sorbonne Universit\'e, 75005 Paris, France.}
\affiliation{Department of Physics, Harvard University, Cambridge, MA 02138, USA}
\author{Joyce Kwan}
\altaffiliation[Current address: ]{JILA, NIST and Department of Physics, University of Colorado, Boulder, Colorado 80309, USA.}
\affiliation{Department of Physics, Harvard University, Cambridge, MA 02138, USA}
\author{Perrin Segura}
\affiliation{Department of Physics, Harvard University, Cambridge, MA 02138, USA}
\author{Yanfei Li}
\affiliation{Department of Physics, Harvard University, Cambridge, MA 02138, USA}
\author{Isaac~Tesfaye}
\affiliation{Institut f\"{u}r Physik und Astronomie, Technische Universit\"{a}t  Berlin, Berlin 10623, Germany}
\author{Gerard Valent\'i-Rojas}
\affiliation{Naquidis Center, Institut d’Optique Graduate School, 33400, Talence, France}
\author{Andr\'e Eckardt}
\affiliation{Institut f\"{u}r Physik und Astronomie, Technische Universit\"{a}t  Berlin, Berlin 10623, Germany}
\author{Markus Greiner}\email{Corresponding author. Email: mgreiner@g.harvard.edu}
\affiliation{Department of Physics, Harvard University, Cambridge, MA 02138, USA}

\date{\today}

%%%%%%% Abstract %%%%%%%

\begin{abstract}
Fractional statistics give rise to quantum behaviors that differ fundamentally from those of bosons and fermions. While two-dimensional anyons play a major role in strongly correlated systems and topological quantum computing, the nature of their one-dimensional (1D) counterparts remains the subject of intense debate, with renewed interest fueled by recent experimental progress. Theoretically, 1D anyons are predicted to host exotic many-body phases and quantum phase transitions, yet experimental signatures have remained elusive. Using ultracold atoms in an optical lattice, we prepare two-body ground states of the 1D anyon-Hubbard model by combining Hamiltonian engineering via quasiperiodic drives and adiabatic state manipulation. We uncover the effects of statistical interactions that lead to pseudo-fermionization and to the formation of chiral bound states when particles remain close together. Our results establish a link between lattice and continuum realizations of anyon models, and mark important steps towards the precise control of 1D anyons in both equilibrium and out-of-equilibrium settings.
\end{abstract}

\maketitle

%%%%%%% Introduction: Adiabatic preparation of 1D anyon ground states %%%%%%%

\begin{bibunit}[apsrev4-2]
    
% \section{Introduction}

The behavior of identical particles is governed by quantum statistics, encoded in the statistical phase $\theta$ that the many-body wave function acquires upon particle exchange. Although limited to bosonic ($\theta = 0$) and fermionic ($\theta = \pi$) statistics in three dimensions, particles called anyons continuously interpolate between these two limits in lower dimensions \cite{1976_Frohlich, 1977_Leinaas, 1982_Wilczek_b, 1988_Frohlich, 2024_Greiter}. Formally, two-dimensional Abelian anyons can be visualized as charge-flux composite particles that acquire an Aharonov-Bohm phase when orbiting each other \cite{1982_Wilczek_a}. Analogously, Abelian anyons in one dimension (1D) \cite{1991_Haldane,2009_Greiter} can be constructed on a 1D lattice using bosons whose tunneling phase depends on density, giving rise to geometric phases that are integer multiples of $\theta$ when particles loop through the configuration space given by lattice-site occupation numbers \cite{2024_Nagies}. During the past decade, the 1D anyon-Hubbard model (AHM) has attracted significant attention due to the novel many-body phases it predicts \cite{2011_Keilmann, 2015_Greschner, 2016_Arcila-Forero, 2017_Zhang, 2024_Bonkhoff} and its connection with topological gauge theories \cite{1995_Rabello, 1999_Kundu, 2021_Bonkhoff, 2022_Frolian, 2022_Chisholm, 2024_Valenti-Rojas}. Recently, a realization of the AHM with ultracold atoms undergoing quantum walks in a driven optical lattice revealed an anyonic Hanbury Brown-Twiss effect manifested as progressive anti-bunching when $\theta$ approaches $\pi$  \cite{2024_Kwan}. In a different setting, the asymmetric momentum distribution of 1D anyons was subsequently observed by injecting a mobile impurity into a strongly interacting quantum gas \cite{2024_Dhar, 2025_Wang}.

\begin{figure}
\includegraphics[width = \columnwidth]{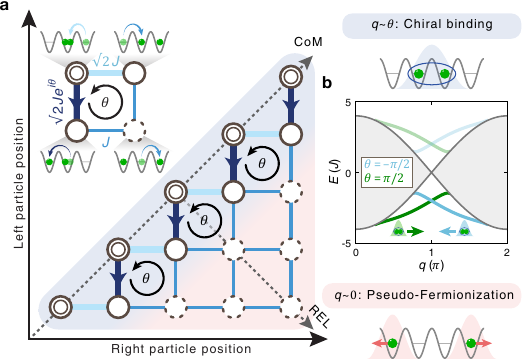} 
\caption{\textbf{Pseudo-fermionization and chiral binding of 1D anyons.} 
(a) Two-particle AHM in position representation.  Due to particle indistinguishability, only the lower right triangle of the coordinate system is relevant. Motion towards the upper right corresponds to shifts of the center of mass (COM), while motion towards the lower right corresponds to increasing the relative distance (REL) between the two particles. Double circles indicating double occupancy. 1D anyons acquire a phase $-\theta$ when tunneling to the right onto an occupied site (dark blue arrow) and $+\theta$ for the reverse process, giving rise to geometric phases around loops in configuration space (inset). The binding of particles requires a non-zero center-of-mass quasi-momentum $q \sim \theta$ due to the tunneling phase, resulting in chiral motion\,\cite{anyons_sm}. The blue-shaded (resp. red-shaded) regions indicate the configurations most populated when the particles are bound (resp. antibunched, i.e. pseudo-fermionized).
(b) Eigenenergies on an infinite lattice, for $U = 0$, with continuum states present for all $\theta$ (grey-shaded region). Bound-state branches, present for $\theta \neq 0$, are chiral when $\theta \neq \pi$ (solid lines for $\theta = \pm \pi/2$). For a system size $L$, the ground state mostly populates states in the lower part of the spectrum with quasi-momenta $q \lesssim \pi/L$.}
\label{fig:theory}
\end{figure}

To illustrate the rich phenomenology of 1D anyons, it is intructive to examine the low-energy properties of two particles described by the AHM. The 1D dynamics of two indistinguishable particles maps onto that of a single particle on a half-plane, as shown in Fig.\,\ref{fig:theory}(a). The states on the diagonal edge then correspond to doubly-occupied lattice sites and the flux $\theta$, piercing only  the plaquettes involving these states and originating from density-dependent tunneling, plays the role of the anyonic exchange phase.  In this representation, the low-energy spectrum naturally separates into bulk-like states, where the two particles remain apart, and edge-like states, corresponding to bound pairs with non-zero center of mass quasimomentum, localized along the diagonal. In the unbound case, the particles show effective anti-bunching induced by destructive interference for non zero values of $\theta$ (see Refs.\,\cite{2016_Strater, 2017_Zhang} and discussion below). The bound state trajectories are expected to be chiral with a preferred direction set by the statistical phase $\theta$ \cite{2016_Cardarelli, 2018_Greschner}, and can be interpreted in the single-particle framework as a chiral edge state induced by the flux $\theta$ along the diagonal. This binding mechanism, which is explained in more detail below, is however different from the topological protection of chiral edge modes in two-dimensional Chern insulators according to the bulk–boundary correspondence. It also lies at the core of the partially-paired superfluid phase introduced in Refs.\,\cite{2016_Cardarelli, 2017_Zhang}. As sketched in Fig.\,\ref{fig:theory}(b), for non-zero $\theta$, bound states appear above and below the continuum of unbound states only for specific values of the center-of-mass quasimomentum $q$, with their chiral nature reflected in the absence of $q \leftrightarrow -q$ symmetry in the dispersion relation. Preparing the ground state of two particles in a reduced system of only three lattice sites, density-dependent tunneling plays a major role and favors a finite quasimomentum which, in turn, provides a large overlap with a bound state in the extended lattice.

In the following, we directly probe the low-energy properties of 1D anyons through adiabatic state manipulations that enable high-fidelity ground state preparation. For the largest system size, we demonstrate pseudo-fermionization of 1D anyons by monitoring the continuous buildup of Friedel oscillations in the density profiles and the evolution of two-particle correlations \cite{2016_Strater}. We further observe asymmetric expansion dynamics signaling the presence of chiral bound states induced by the statistical phase, and confirm the chiral nature of these bound states by measuring their reflection dynamics against a potential barrier.

Our experiments start with ultracold $^{87}$Rb atoms loaded in a two-dimensional optical lattice at the focus of a high-resolution imaging system. Particle motion is restricted to one spatial dimension $x$ by maintaining a deep lattice potential along the transverse direction. We induce a density-dependent Peierls phase by modulating the lattice depth $V_x$ with three frequency components in the presence of a strong potential gradient \cite{2016_Cardarelli, 2024_Kwan, anyons_sm}, realizing a generalized Bose-Hubbard model
\begin{multline}
\label{eq:H_eff}
\hat{H} = - J \sum_j \left( \hat{b}_{j+1}^{\dagger} \e^{-\mi \hat{n}_{j+1} \theta} \hat{b}_j + \mathrm{h.c.}\right) \\ 
+ \frac{U}{2} \sum_j \hat{n}_j \left( \hat{n}_j - 1 \right) + \Delta \sum_j j \, \hat{n}_j,
\end{multline}
where $J$ is the tunneling amplitude, $U$ the on-site interaction energy, $\Delta$ the energy offset between adjacent lattice sites, which parametrizes the effective tilt along the chain, and $\hat{b}_j^{\dagger}$, $\hat{b}_j$ and $\hat{n}_j$ are respectively the bosonic creation, annihilation, and number operators at site $j$. Most importantly, $-\hat{n}_j \theta$ ($\theta \in [-\pi, +\pi]$) is the density-dependent Peierls phase acquired upon tunneling right from site $j-1$ to site $j$, which effectively imprints anyonic statistics in the dynamics. For state initialization, we use digital micromirror devices (DMDs) to isolate two atoms on the same site within the doubly-occupied region of a Mott insulator. A DMD is then used to confine the dynamics to a finite chain of $L$ sites.  For $\Delta>U>0$ and in the absence of tunneling ($J = 0$), the initial configuration $|2,0,0,...\rangle$, corresponding to a doublon at the bottom of the chain, is the ground state of the system. This state is then adiabatically transferred to the ground state of the $L$-site chain [Fig.\,\ref{fig:friedel}(a)]. At the end of the evolution, we extract full-counting statistics of the quantum state using fluorescence imaging and post-select only snapshots containing exactly two particles \cite{anyons_sm}.

%%%%%%% Section 1: Observation of Friedel oscillations in the density profiles %%%%%%%

\section{Buildup of Friedel oscillations}

\begin{figure*}
\includegraphics[width = \textwidth]{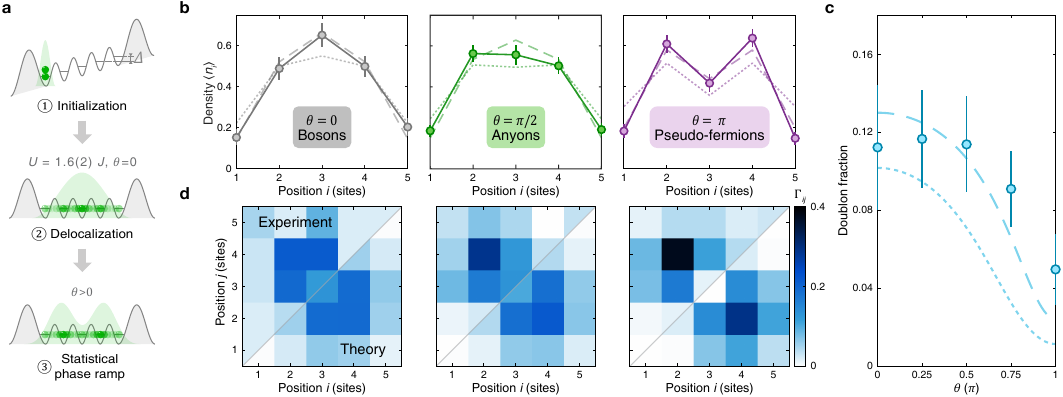} 
\caption{\textbf{Friedel oscillations and pseudo-fermionization.} 
(a) Adiabatic preparation of the two-particle ground state on a chain of $L = 5$ lattice sites with on-site interaction $U = 1.6(2)J$. Starting from a doubly-occupied site at the bottom of the chain (\raisebox{.5pt}{\textcircled{\raisebox{-.9pt} {1}}}), tunneling is increased to $J$, followed by a decrease of the tilt $\Delta$ from $4.3(2)J$ to $0$ (\raisebox{.5pt}{\textcircled{\raisebox{-.9pt} {2}}}). Finally, the statistical phase is ramped from $0$ to its final value $\theta$ (\raisebox{.5pt}{\textcircled{\raisebox{-.9pt} {3}}}).
(b) Average density profiles evolve from a single central peak at $\theta = 0$ (bosons) to two side peaks at $\theta = \pi$ (pseudo-fermions).
(c) The fraction of doubly-occupied states decreases as $\theta$ approaches $\pi$.
(d) Density-density correlations $\Gamma_{ij}$, representing the probability of finding particles at $i$ and $j$, confirm the increasing tendency of particles to avoid one another.
Theory predictions, obtained from exact diagonalization \cite{anyons_sm}, are shown as dotted curves in (b) and (c) for parameter-free calculations, while dashed curves include adjusted offset potentials. In (d), only theory values with edge offsets are shown. Error bars represent the standard error of the mean (s.e.m.).
}
\label{fig:friedel}
\end{figure*}

We first investigate the emergence of Friedel oscillations, a hallmark of fermionization, as the statistical parameter $\theta$ is increased from $0$ to $\pi$ in the minimal setting of two particles \cite{2016_Strater}. To this end, we prepare two-particle ground states on a finite system of $L = 5$ lattice sites for various statistical phases $\theta$. High-fidelity adiabatic preparation is best achieved by keeping the excitation gap as large as possible during the adiabatic passage \cite{2023_Leonard, anyons_sm}. In practice, we first delocalize the two particles by increasing tunneling to its final value $J/h = 10.4(2)$ Hz within 50\,ms, where $h$ is Planck's constant, while the tilt $\Delta = 4.3(2)J$, the repulsive on-site interaction energy $U = 1.6(2)J$, and the statistical phase $\theta = 0$ are kept constant [Fig.\,\ref{fig:friedel}(a)]. We then lower the tilt $\Delta$ over 250\,ms, before ramping up $\theta$ to its final value \cite{anyons_sm}. By reversing the ramp protocol and after post-selection, we measure a probability greater than $80\%$ of recovering the initial doubly-occupied site for all data presented below, confirming good adiabaticity \cite{anyons_sm}.

The ground-state density profiles of two particles on a chain of $L = 5$ sites are shown in Fig.\,\ref{fig:friedel}(b) for different statistical phases at the end of the adiabatic ramp. Dotted lines correspond to parameter-free theoretical predictions from exact diagonalization, while the dashed lines show the same theory with edge potentials fitted to the measured density profiles, accounting for residual confinement offsets at the boundaries \cite{anyons_sm}. For bosonic particles ($\theta = 0$), we observe a single density peak centered on the middle site. As the statistical phase $\theta$ increases, the density profile gradually flattens until a density dip is formed for $\theta = \pi$, with the two density peaks suggesting that the particles tend to avoid each other. We interpret this density dip as a minimal version of so-called Friedel oscillations, a hallmark of fermionic behavior that arises from localized perturbations (like the sharp walls at the edges of our system) in metallic or semiconductor systems \cite{1958_Friedel, 2003_Giamarchi, 2016_Strater, 2025_Emperauger}. Strictly speaking, particles with statistical phase $\theta = \pi$ are pseudo-fermions as they do not obey the Pauli principle, acting as fermions when located on different sites and as bosons when occupying the same site \cite{2024_Kwan}. While the anyon model here is soft-core, meaning there is no intrinsic Pauli principle, pseudo-fermions behave very similarly to true fermions as the system becomes more dilute \cite{2016_Strater}. This behavior can be understood as the result of destructive interference between the two possible tunneling processes that create a doubly occupied site, with the second particle tunneling from the left with phase $-\theta$ [dark blue line in Fig.\,\ref{fig:theory}(a)] or from the right with phase $0$ [light blue line in Fig.\,\ref{fig:theory}(a)].

Full counting statistics allows us to confirm this interpretation by giving access to two-body correlations. The doublon fraction, shown in Fig.\,\ref{fig:friedel}(c), decreases with increasing $\theta$, reflecting the growing influence of statistical repulsion. More generally, we show the density-density correlator $\Gamma_{ij} = \langle b_j^\dagger b_i^\dagger b_i b_j \rangle$ between sites $i$ and $j$ in Fig.\,\ref{fig:friedel}(d) \cite{doublon}. Although a tendency for the two particles to occupy different sites is already present at $\theta = 0$ due to repulsive on-site interactions $U > 0$, the correlations show a significant enhancement of this tendency as $\theta \rightarrow \pi$. As originally proposed in Ref.\,\cite{2016_Strater}, our observations provide clear evidence of the continuous pseudo-fermionization of 1D anyons as the statistical phase $\theta$ is tuned from $0$ to $\pi$. In particular, pseudo-fermionization originates from anyonic statistics rather than from the explicit repulsive on-site interactions $U$ introduced in our experiment to maintain a sufficiently large excitation gap for adiabatic state preparation.

%%%%%%% Section 2: Expansion dynamics %%%%%%%

\section{Asymmetric expansion dynamics}

\begin{figure}
\includegraphics[width = \columnwidth]{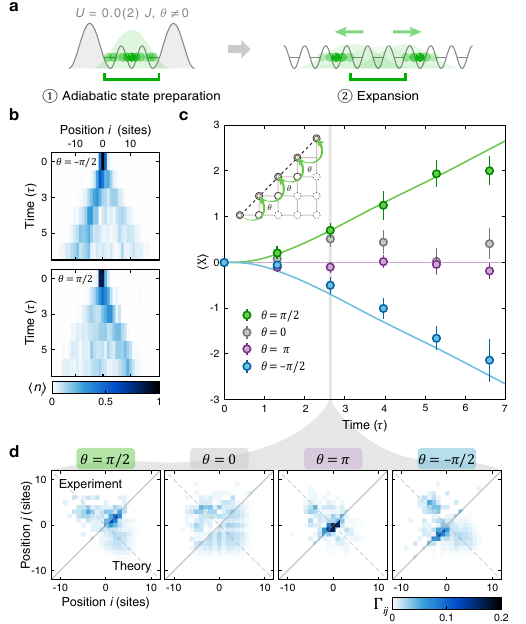} 
\caption{\textbf{Asymmetric expansion dynamics.}
(a) Adiabatic preparation of the two-particle ground state on a chain of $L = 3$ lattice sites, with $U = 0.0(2)J$: after isolating a doubly-occupied site at the bottom of the chain, tunneling is increased to $J$, before the tilt is decreased to $0$ at constant $\theta$ (\raisebox{.5pt}{\textcircled{\raisebox{-.9pt} {1}}}). The confining walls are then quenched to zero to allow free expansion of the system (\raisebox{.5pt}{\textcircled{\raisebox{-.9pt} {2}}}). 
(b) Density profile after expansion for $\theta = -\pi/2$ (resp. $\theta = +\pi/2$) shows leftward (resp. rightward) trajectory.
(c) Center-of-mass position $\langle X \rangle$ as a function of time for different statistical phases. For $\theta \neq 0$ or $\pi$, the observed drift direction matches the one predicted for the bound pair (inset). 
(d) Density-density correlator $\Gamma_{ij}$ at time $t = 3.6\,\tau$, showing dominant correlations near the diagonal, consistent with chiral bound states induced by the statistical phase $\theta$.
Theory predictions are based on exact diagonalization without free parameter \cite{anyons_sm}.
}
\label{fig:asymmetric}
\end{figure}

In order to study the bound states and their chiral nature, we prepare the ground state of two particles on three lattice sites only. In this small system, three out of six occupation number states that span the Hilbert space are doubly occupied, such that the density-dependent tunneling phases characterized by $\theta$ determine the structure of the ground state and lead to a non-zero center-of-mass quasimomentum. As a result, this state has a large overlap with a bound pair on the full chain, extending beyond three sites, allowing the bound pair to be studied through expansion dynamics. Smaller systems provide larger excitation gaps, which simplifies the preparation protocol: the two particles are delocalized by increasing tunneling to $J$, followed by a reduction of the tilt $\Delta$ from $4.3(2)J$ to 0, while keeping the on-site interaction [$U = 0.0(2)J$] and the statistical phase $\theta$ constant. We then quench the optical confinement to let the system expand freely under the AHM [Fig.\,\ref{fig:asymmetric}(a)], and the system is then detected after a variable expansion time $t$, expressed in units of the inverse tunneling rate $\tau = h/(2 \pi J) = 15.3(3)$ ms.

The expansion dynamics of a pair of 1D anyons initially prepared on $L = 3$ sites are shown in Fig.\,\ref{fig:asymmetric}(b) for $\theta = \pm \pi/2$. Although the particles are initially at rest, the cone-shaped density profiles reveal a ballistic expansion with a preferred direction of motion that depends on the sign of $\theta$ \cite{anyons_sm}. This directional expansion is confirmed by the time evolution of the center-of-mass position $\langle X \rangle$ [Fig.\,\ref{fig:asymmetric}(c)]. In contrast, the center of mass remains stationary  for bosons ($\theta = 0$) and pseudo-fermions ($\theta = \pi$). A short-time perturbative expansion of the initial state reveals that, to leading order, $\langle X\rangle = \sin(\theta) / [9(\sqrt{5+4\cos(\theta)}+5)]^{1/2} \, (t/\tau)^{3}$, demonstrating that the asymmetric expansion has a cubic onset in time\,\cite{anyons_sm}. Such asymmetric expansion originates from the simultaneous breaking of parity and time-reversal symmetries in the AHM for $\theta \neq 0$, $\pi$ \cite{2018_Liu}. Interpreting the density-dependent Peierls phase as a dynamical gauge potential proportional to the local density, the outward motion of the cloud causes the gauge potential to evolve in time, thereby generating an electric force on the particles and inducing a finite drift velocity \cite{anyons_sm, 2022_Frolian, 2022_Chisholm}. We further study these dynamics by analyzing the density-density correlator $\Gamma_{ij}$ at a fixed evolution time $t = 3.6\,\tau$ [Fig.\,\ref{fig:asymmetric}(d)]. Most of the weight contributing to the observed asymmetry lies near the diagonal, corresponding to particles that remain close together. These strong diagonal correlations indicate the presence of bound states \cite{2024_Kwan}. 

By using adiabatic state preparation, our protocol selectively populates the lower bound-state branch of the two-particle spectrum [Fig.\,~\ref{fig:theory}(b)]. Namely, the large density in the initial three-site ground state enhances the influence of the Peierls phases, which shifts the total mean quasimomentum to a finite value, making bound states in the enlarged post-quench system energetically favorable. This lower branch exhibits pronounced chirality, with a predominantly negative (resp. positive) group velocity $v_g = \dd E / \dd q$ for $\theta = \pi/2$ (resp. $\theta = -\pi/2$), in agreement with the trajectories shown in Fig.\,\ref{fig:asymmetric}(c). The origin of the bound states can be understood from the single-particle representation shown in Fig.\,\ref{fig:theory}(a): motion along the diagonal edge corresponds to center-of-mass motion, where density-dependent Peierls phases shift the dispersion. As a result, the energy minimum occurs at finite quasimomentum $q \sim \theta$, while for distant particles, where no Peierls phases are present, the minimum lies at $q=0$. This effect, together with the bosonic enhancement of tunneling along the diagonal edge, lowers the kinetic energy for configurations near the diagonal and provides a momentum-dependent binding mechanism \cite{anyons_sm}. Note that this effect is different from the topological protection of chiral edge states at the boundary of Chern insulators. Note also that in Fig.\,\ref{fig:asymmetric}(b,d), we attribute the lack of symmetry between the experimental profiles for $\theta$ and $-\theta$ to potential disorder along the 1D chain \cite{2018_Liu, anyons_sm}.

\section{Probing the dynamical binding mechanism}

\begin{figure}
\includegraphics[width = 1\columnwidth]{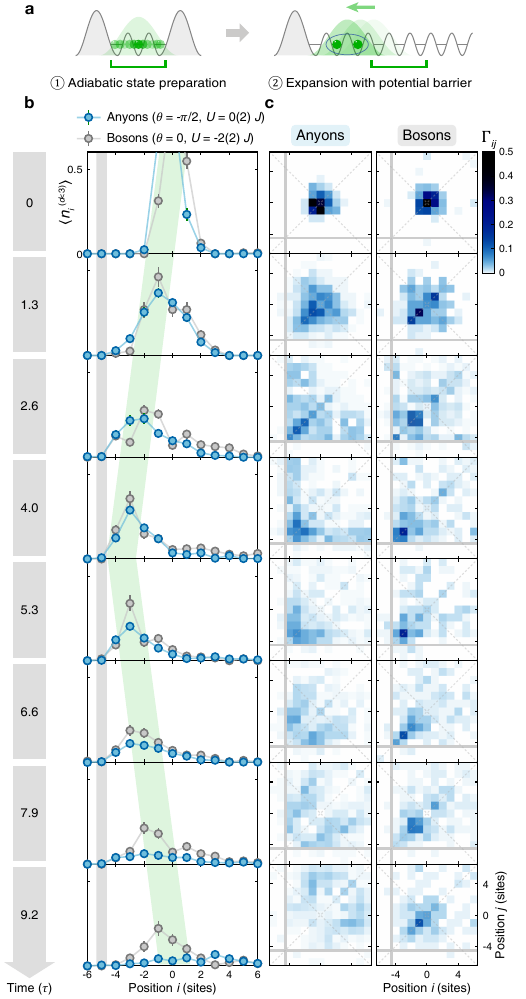} 
\caption{\textbf{Reflection dynamics of two-particle bound states.}
(a) Following the preparation steps described in Fig.~\ref{fig:asymmetric}(a) (\raisebox{.5pt}{\textcircled{\raisebox{-.9pt} {1}}}), the confining walls are quenched to zero to allow expansion in the presence of a potential barrier located four sites to the left of the initial position (\raisebox{.5pt}{\textcircled{\raisebox{-.9pt} {2}}}). Attractive bosons are directed towards the potential barrier through a controlled momentum transfer at $t = 0$ \cite{anyons_sm}.
(b) Density profiles conditioned on particle pairs with relative distance $d < 3$ sites. At early times, both anyons ($\theta = -\pi/2$, $U = 0.0(2)J$, blue dots) and attractive bosons ($\theta = 0$, $U = -2.0(2) J$, grey dots) exhibit similar leftward motion. After reflection around $t = 4\tau$, the two cases show qualitatively different behavior: the density of nearby anyons decreases significantly, while bosons remain tightly bound. The green-shaded area indicates the approximate trajectory of the center of mass.
(c) Density-density correlators $\Gamma_{ij}$ reveal that attractive bosons reflect while staying together (strong weight along the diagonal), whereas chiral anyons delocalize after reflection, revealing the chiral nature of the binding mechanism.}
\label{fig:wall}
\end{figure}

To further probe the chirality of the bound states induced by the statistical phase $\theta$, we measure their reflection dynamics against a potential barrier \cite{2022_Frolian}. Specifically, we use adiabatic state preparation to preferentially populate a two-body bound state with a well-defined chirality \cite{anyons_sm}. As in the previous case, we prepare the ground state of two particles on $L=3$ lattice sites with $\theta=-\pi/2$ and $U=0.0(2)$, and then let the system expand in the presence of a potential barrier positioned four sites to the left of the initial sites [Fig.\,\ref{fig:wall}(a)]. For comparison, we also investigate the reflection dynamics of a non-chiral bound state. In this case, we prepare two particles in the ground state at $\theta = 0$ and $U = -2.0(2) J$, populating a non-chiral attractive bound-state branch. In this case, the atoms are directed towards the potential barrier through a controlled momentum transfer \cite{anyons_sm}.

To enhance sensitivity to differences in bound-state dynamics, we restrict our analysis to particle pairs with relative distance of $d < 3$ sites, conditioning the density profiles accordingly [Fig.\,\ref{fig:wall}(b)]. At short times, the dynamics of nearby particles are similar in both cases, showing a net motion to the left. However, the wave packets start to differ significantly after reflecting off the barrier around $t = 4\tau$. In particular, the integrated density of nearby anyons [$\theta = -\pi/2$, $U = 0.0(2)$] decreases significantly, whereas attractive bosons [$\theta = 0$, $U = - 2.0(2)J$] exhibit nearly elastic reflection. This contrast is also visible in the density-density correlator, which captures correlations across all separations $d$ [Fig.\,\ref{fig:wall} (c)]. While attractive bosons retain proximity after reflection, anyons cannot do so while moving rightwards due to the chiral nature of the binding mechanism, which favors bound states only for particles moving to the left for positive $\theta$ [Fig.\,\ref{fig:theory}(b)]. We note that the measured expansion dynamics deviates from parameter-free theory predictions, likely due to residual potential gradients \cite{anyons_sm}. These experiments provide direct evidence for the chiral nature of 1D anyon bound states in the strongly correlated regime, and establish a connection with the chiral solitons realized in Ref.\,\cite{2022_Frolian} using an optically-dressed Bose-Einstein condensate \cite{1996_Aglietti, 2022_Chisholm}. Although beyond the scope of the present work, bridging the gap between the two-body quantum regime explored here and the classical field limit explored in Refs.\,\cite{2022_Frolian, 2022_Chisholm} remains an important open question.

\section{Conclusion and perspectives}

Using adiabatic state preparation, we have observed two complementary manifestations of one-dimensional anyonic statistics: continuous pseudo-fermionization, revealed by Friedel oscillations and suppressed two-body coincidences, and the emergence of chiral bound states. Our approach to preparing two-body ground states provides access to a broad range of physical phenomena, including chirally protected zero-energy states \cite{2025_Theel} and bound states in the continuum, which are spatially localized eigenstates embedded within a continuum of scattering states \cite{2016_Hsu}. It also enables the study of bound state collisions by preparing two spatially separated bound states with opposite chirality \cite{2024_Su}.

Beyond the two-body setting, the ability to reliably prepare few-body ground states can be extended to larger systems, enabling the realization of anyon-cluster states and the investigation of their connection to chiral solitons \cite{1998_Griguolo, 2017_Posske, 2022_Frolian, 2022_Chisholm}, as well as the emergence of novel superfluid phases \cite{2011_Keilmann, 2015_Greschner, 2016_Arcila-Forero, 2017_Zhang}. Progress in this direction would benefit from advanced optimization techniques for adiabatic protocols already developed in the context of quantum gas microscopy \cite{2024_Blatz}, and extendable to periodically-driven quantum systems \cite{2024_Schindler, 2025_Schindler}. Finally, density-dependent tunneling provides a versatile platform to simulate exotic forms of magnetism in ladder geometries \cite{2015_Greschner_b, 2016_Mishra}, to engineer flux attachment in two dimensions \cite{2019_Barbiero, 2020_Valenti-Rojas}, and to realize so-called traid anyons, enabled by three-body hardcore interactions, in one-dimensional systems \cite{2024_Nagies}.

We thank R\'emy Vatr\'e and Botao Wang for valuable comments on the manuscript, and Marin Bukov, Alessio Celi, Nathan Goldman, Nathan Harshman, Lukas Homeier, Christina Sonja Mascherbauer, Thore Posske, Luis Santos, Leticia Tarruell, Amit Vashisht, R\'emy Vatr\'e and Botao Wang for valuable discussions. We acknowledge support by the ONR (award No.~N00014-18-1-2863), LBNL/DoE QSA (award No.~DE-AC02-05CH11231), ARO (award No.~W911NF-20-1-0021), Gordon and Betty Moore Foundation (award
No.~GBMF-11521), and NSF (award Nos.~PHY-1734011, PHY-2317134, and OAC-2118310). A.E. and I.T. were supported by the Deutsche Forschungsgemeinschaft (DFG, German Research Foundation) via the Research Unit FOR 5688 under project number 521530974. I.T. acknowledges support from the Studienstiftung des deutschen Volkes. G.V.-R. acknowledges financial support from the Naquidis Center for Quantum Technologies.

M.G. is a cofounder and shareholder of QuEra Computing. All other authors declare no competing interests.

%%%%%%% Bibliography %%%%%%%

\putbib % inserts references for this unit
\end{bibunit}

%%%%%%% Supplements %%%%%%%
\newpage \clearpage\hbox{}\thispagestyle{empty}

%%%%%%%%%% Merge with supplemental materials %%%%%%%%%%
%%%%%%%%%% Prefix a "S" to all equations, figures, tables and reset the counter %%%%%%%%%%
\setcounter{equation}{0}
\setcounter{figure}{0}
\setcounter{table}{0}
\setcounter{page}{1}
\makeatletter
\renewcommand{\theequation}{S\arabic{equation}}
\renewcommand{\thefigure}{S\arabic{figure}}
\nocite{*}
\renewcommand{\bibnumfmt}[1]{[S#1]}
\renewcommand{\citenumfont}[1]{S#1}
%%%%%%%%%% Prefix a "S" to all equations, figures, tables and reset the counter %%%%%%%%%%

\begin{center}
\textbf{\large Supplemental Material}
\end{center}

\begin{bibunit}[apsrev4-2]

\section{Experimental sequence and calibrations}

\subsection{Initial state preparation}

The experiments reported in this work start by preparing a Bose-Einstein condensate of ${}^{87}\mathrm{Rb}$ atoms in the $\left|F=1, m_F=-1\right\rangle$ hyperfine state. The atoms are loaded into a single dark fringe of a one-dimensional blue-detuned optical lattice along the vertical direction, with a lattice spacing of $a_z = 1.5\,\si{\micro\meter}$ and a lattice depth $V_z = 250 E_R^z$, where the recoil energy is defined as $E_R^z = \hbar^2 / (2ma_z^2) = h \times 0.25\,\si{kHz}$ and $m$ the mass of a ${}^{87}\mathrm{Rb}$ atom. A Mott insulator is then obtained by ramping up a two-dimensional optical lattice in the $(x,y)$ plane with lattice spacing $a = 680\,\si{nm}$ and depths $V_x = V_y = 45 E_R$, where
$E_R = \hbar^2 / (2ma^2) = h \times 1.24\,\si{kHz}$.

Two independent copies of our initial state, each consisting of two atoms on the same site, are prepared by optically confining two adjacent doubly-occupied sites along the $y$ direction within the $n=2$ region of the Mott insulator. The remaining atoms are expelled using a repulsive optical potential with a large waist. The confining potential used to isolate the doublons is realized by holographically projecting repulsive potentials using two digital micromirror devices (DMDs) placed in the Fourier plane of our high-resolution imaging system. One of these DMDs is then used to project two sharp repulsive walls separated by $L + 1$ sites along the $x$ direction, realizing a finite chain of $L$ sites, with the doublons initially positioned at one edge of the accessible region.

\subsection{Hubbard parameters}

\begin{figure}
\includegraphics[width = 1.0\columnwidth]{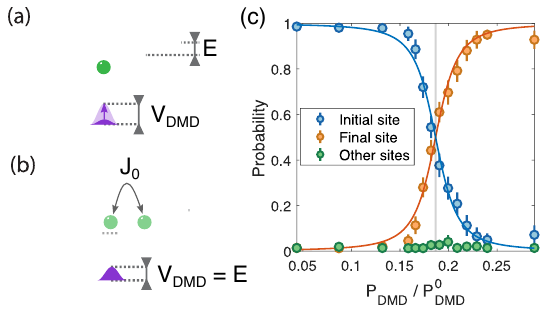} 
\caption{\textbf{Calibration of DMD wall potential.} 
(a) Tilted lattice along $x$ together with the additional local potential created by the DMD (purple).
(b) The DMD potential is increased to the resonance condition $V_{\textrm{DMD}} = E$, enabling resonant tunneling between neighboring sites. 
(c) Measured probability of finding the atom on the initial site (blue), on the neighboring site to the right (orange), or on other sites (green), as a function of the applied DMD laser power. The resonance condition $V_{\textrm{DMD}} = E$ is extracted from the transfer probability to the neighboring site.
}
\label{Fig_SM:DMD}
\end{figure}

The experiments are performed by maintaining a large lattice depth $V_y = 45 E_R$ along the $y$ direction in order to decouple the different tubes oriented along the $x$ direction. A magnetic field gradient along $x$ is ramped up to induce an offset energy $E = h \times 801(2)$\,Hz between adjacent lattice sites. The lattice depth $V_x$ is then ramped down to $6 E_R$, yielding a bare tunneling amplitude $J_0 = h \times 79(2)$\,Hz, calibrated via 1D quantum walks in a flat lattice, as described in Ref.\,\cite{2024_Kwan}. The bare on-site interaction energy $U_0 = h \times 192(2)$\,Hz is extracted, together with the tilt $E$, from lattice amplitude modulation spectroscopy \cite{2024_Kwan}.

In the absence of lattice depth modulation, atoms remain Stark localized due to the large tilt $E \gg J_0, U_0$. Tunneling is restored with a density-dependent Peierls phase by modulating the lattice depth $V_x$ simultaneously with three frequency components $\{ \omega_{-1}, \omega_0, \omega_{+1} \} = \{ E + \Delta - (U_0 - U),\, E + \Delta,\, E + (U_0 - U) \}/\hbar$, each with the same relative amplitude $\delta V (\omega) / V_x$. This realizes the effective Hamiltonian~\eqref{eq:H_eff} described in the main text, with tunable tunneling amplitude $J$, on-site interaction $U$, and residual tilt $\Delta$. In particular, a maximum effective tunneling amplitude $J = h \times 10.2(2)$\,Hz is realized with a modulation amplitude $\delta V (\omega) / V_x = 20\,\%$ \cite{2024_Kwan}. 

The offset energy induced on the outer sites by the DMD wall potential is calibrated by inducing resonant tunneling between two adjacent lattice sites in the presence of the tilt $E$. To do so, we initialize the system with a single atom in the chain under tilt, see Figure\,\ref{Fig_SM:DMD}(a). With the DMD potential centered on this occupied site, we adiabatically increase the DMD laser power over 100 ms, then measure the final location of the atom, as shown in Figure\,\ref{Fig_SM:DMD}(c). Fitting the transfer probability to the next site assuming perfect adiabaticity yields a resonance condition where the energy difference between the two sites is $\delta V_{\textrm{DMD}} = E$ [Figure\,\ref{Fig_SM:DMD}(b)]. This value can then used to extract the outer site energy offset $V_{\textrm{DMD}} = 4(1)\,\si{\kilo\hertz}$ by rescaling with the increase in laser power between the calibration measurement at resonance ($P_\textrm{DMD}$) and the laser power used in the experimental sequence ($P_\textrm{DMD}^0$).

\subsection{Detection and post-selection}

\begin{figure}
\includegraphics[width = \columnwidth]{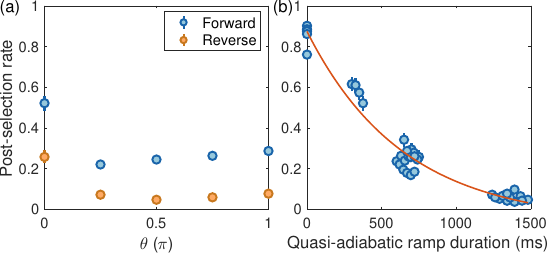} 
\caption{\textbf{Post-selection rate.}
(a) Post-selection rates for the dataset shown in Figure\,\ref{fig:friedel}, for the forward (blue dots) and reversed (orange dots) used to estimate adiabaticity, as a function of the final statistical phase $\theta$.
(b) Post-selection rates for an extended dataset as a function of the total quasi-adiabatic ramp duration, fitted to an exponential decay yielding a single-particle lifetime $\tau = 0.6(1)$\,s.
}
\label{Fig_SM:post-selection}
\end{figure}

We detect the final quantum state in the Fock basis by performing fluorescence imaging of the atomic distribution, after switching to a deep optical lattice at $795\,\si{\nano\meter}$. To avoid losing pairs of atoms due to light-assisted collisions, we expand the atoms along the $y$ direction beforehand by lowering the corresponding lattice depth, while separating the two independent chains using a sharp wall potential along $x$, as described in Ref.\,\cite{2016_Kaufman}.

In all data shown in the main text, we post-select experimental snapshots that contain exactly two atoms. As shown in Figure\,\ref{Fig_SM:post-selection}(a), which complements the data in Fig.\,\ref{fig:friedel} of the main text, post-selection rates are below $30\%$ after the forward adiabatic ramp when including a $\theta$ ramp, and below $10\%$ after the reverse adiabatic ramp used to estimate the overall adiabaticity. Figure\,\ref{Fig_SM:post-selection}(b) shows the post-selection rate as a function of the quasi-adiabatic ramp duration for an extended dataset, yielding a single-particle lifetime of $\tau = 0.6(1)\,\si{\second}$. While a comprehensive discussion of the loss mechanisms is beyond the scope of this work, we attribute the low post-selection rates to interband Floquet heating that leads to atom loss, while the relatively large return probability achieved here suggests that intraband heating remains comparatively weak\,\cite{2020_Sun}.

\subsection{Numerical simulations}

All numerical predictions were obtained via exact diagonalization and real-time evolution using the effective Hamiltonian \eqref{eq:H_eff}, with parameters chosen according to our experimental calibrations. To account for the observed reduction of density at the edges in Fig.\,\ref{fig:friedel}, the edge potentials at the boundaries were treated as a free parameter and fitted using a chi-square minimization across all density profiles, yielding an offset potential $V_{\textrm{edge}} = 0.65\,J = 6.6(1)$\,Hz, consistent with residual offsets induced by the DMD confining light potential.

\subsection{Quasi-adiabatic ramps}

\begin{figure}
\includegraphics[width = \columnwidth]{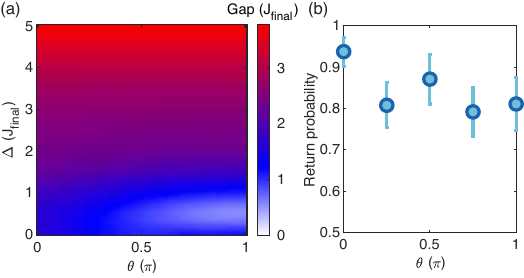} 
\caption{\textbf{Quasi-adiabatic ramps.}
(a) Gap to first excited state for a two-particle on 5-site system as a function of the residual tilt $\Delta$ and statistical parameter $\theta$. For the ground states in Figure\,\ref{fig:friedel}, the ramp pathway we take is to first lower the $\Delta$ at $\theta=0$, and then ramp $\theta$ to the target value.
(b) The return probability, a metric of adiabaticity, for the ground states realized in Figure\,\ref{fig:friedel}.
}
\label{Fig_SM:quasi-adiabatic}
\end{figure}

To realize the ground states in Fig.\,\ref{fig:friedel}, we adiabatically connect the initial state $|2,0,0,...\rangle$ to the target state, taking a ramp pathway that is informed by the gap between the ground state and the first excited state. As shown in Fig.\,\ref{Fig_SM:quasi-adiabatic}(a), after increasing tunneling $J$ to its final value, a possible pathway is to first lower the tilt $\Delta$ to 0 at $\theta = 0$ and then increase $\theta$ from $0$ to the final value. We take this pathway and optimize the ramp time by maximizing the return fidelity, which is a metric for adiabaticity and is the probability the system has returned to the initial state $|2,0,0,...\rangle$ after ramping to the target state and then reversing the ramp back to the initial parameters. 

We optimize the ramp in a piecewise fashion. First, we determine the optimal time to lower the tilt $\Delta$ to zero. Fixing this optimal segment, we then scan the time to increase $\theta$ from $0$ to the final value. Fig.\,\ref{Fig_SM:quasi-adiabatic}b shows the maximum return fidelities obtained for target ground states at various $\theta$. Assuming that the forward evolution to the target state is independent of the reverse ramp back to the initial parameters, we can estimate the overlap with the target state by taking the square root of the measured return fidelity. %, which is also shown in Fig.\,\ref{Fig_SM:quasi-adiabatic}b.

Realizing the two-particle ground states on three sites in Figs.\,\ref{fig:asymmetric} and \ref{fig:wall} is more straightforward because the reduced system size leads to larger excitation gaps and allows for a simpler ramp. Instead of the two-part ramp for the larger system, where we first lower $\Delta$ and then increase $\theta$, we directly lower $\Delta$ at the $\theta$ of interest to ramp to the target state. When optimizing this ramp, we obtain a return probability $\sim 80 \%$ for each $\theta$.

\subsection{Momentum kick for bosons}

\begin{figure}
\includegraphics[width = \columnwidth]{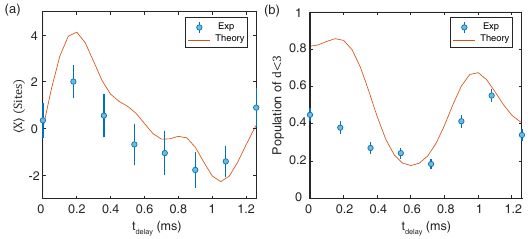} 
\caption{\textbf{Momentum transfer of bound states of bosons.}
(a) Center-of-mass position $\langle X \rangle$ at an expansion time $t = 5.2 \tau$ for attractive bosons as a function of of $t_{\textrm{delay}}$. The center of mass is shifted due to the phase kick, the amount of momentum transfer being controlled by $t_{\textrm{delay}}$.
(b) Population of particle pairs with relative distance $(d < 3)$ as a function of $t_{\textrm{delay}}$.
}
\label{Fig_SM:Boson_com_drift}
\end{figure}

To compare the reflection behaviors of anyonic and bosonic bound states in Fig.\,\ref{fig:wall}, we match their initial center-of-mass velocity. While anyonic bound states are naturally chiral, bosonic bound states do not have a preferred direction of motion unless kicked. To achieve this, we implement a controlled momentum transfer method: we quench the phases of the three-tone drive to launch the particles at the desired velocity. We first prepare the ground state of the attractive bosons at $t = 0$, then quench the drive phases to $(\phi_{-1}, \phi_{0}, \phi_{+1}) = (\omega_{-1}, \omega_{0}, \omega_{+1} ) t_{\textrm{delay}}$ and simultaneously turn off the confining walls. Fig.\,\ref{Fig_SM:Boson_com_drift}(a) shows the center-of-mass drift after an expansion time $t = 5.2 \tau$ as a function of the time delay $t_{\textrm{delay}}$. Experimentally, we find that $t_{\textrm{delay}} = 1.08$\,ms reproduces approximately the center-of-mass velocity measured for [$\theta = -\pi/2$, $U = 0.0(2)J$] anyons, while keeping a large fraction of nearby atoms and therefore not breaking up the bound pair [Fig.\,\ref{Fig_SM:Boson_com_drift}(b)].

\subsection{Residual tilt in the expansion dynamics}

Our numerical simulations support qualitatively the conclusions drawn from Fig.\,\ref{fig:wall} of the main text: when reflected off a potential barrier, the anyonic bound state breaks apart, while the attractively-bound state of bosons reverses direction and remains bound. However, direct comparison between theory and experiment is challenging due to the sensitivity of the anyon bound state propagation to the Hamiltonian parameters. In particular, the timescale for the breaking of the anyonic bound pair in the experiment is faster than in the numerical results. Quantitatively, by fitting the tunneling amplitude to our experimental results, we obtain a value ($J_{\textrm{fit}} = 1.30(1)\,J$) significantly outside the uncertainty interval ($\delta J = 0.02\,J$) obtained from independent quantum walk calibrations.

\begin{figure}
\includegraphics[width = \columnwidth]{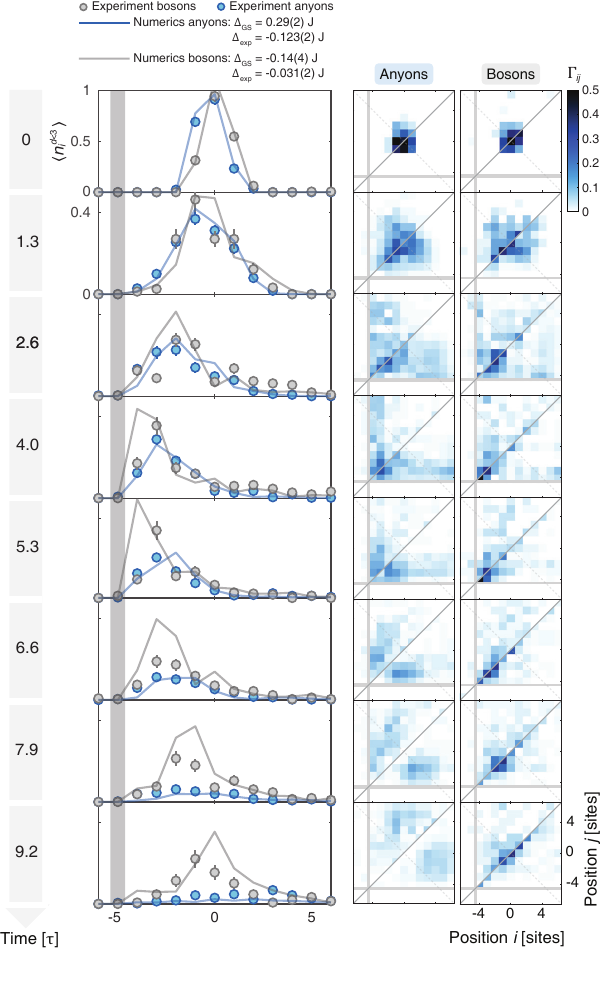} 
\caption{\textbf{Bound state expansion compared with simulations}
Left: density of nearby particles during expansion for anyons and bosons, with experimental data (dots) and fitted theoretical curves (lines).
Right: two-particle correlations compared with the fitted theoretical predictions.
}
\label{Fig_SM:F4_tiltFit}
\end{figure}

The presence of a residual potential gradient can have a similar effect on the bound state dynamics. We therefore fit this residual gradient to our experimental data. More precisely, we extract two separate gradient values, one before and one during expansion, to account for a possible additional gradient introduced by the walls during the preparation of the ground state. After fitting the initial gradient to the measured density profile at $t = 0$, we then use the resulting ground state as the starting point to fit an expansion gradient to the density profiles obtained during expansion. The results in Fig.\,\ref{Fig_SM:F4_tiltFit} show that a ground state tilt of $0.29(2)\,J$ and an expansion gradient of $-0.12(1)\,J$ accurately reproduce the measured density profiles. Since the anyonic and bosonic expansion data were taken two weeks apart, it is reasonable to assume that the potential landscape may have drifted, so we fit the bosonic data separately. These results are also plotted in Fig. \ref{Fig_SM:F4_tiltFit}, with an optimal ground state gradient equal to $-0.14(4)~J$ and an expansion gradient of $-0.03(1)~J$. In particular, the fitted gradients lie within the experimental error bars. Regardless of the specific structure of the potential landscape, scattering with a small potential defect will break apart the anyonic bound state while leaving the bosonic bound state intact.

\section{Generalities on the Anyon-Hubbard model}

\subsection{Boson--anyon mapping}

A standard way to introduce one-dimensional anyons on a lattice is via a generalized Jordan--Wigner transformation that maps local bosonic operators $b_j$ to anyonic operators $a_j$, defined by
\begin{equation}\label{eq:jwtrans}
a_j = b_j \exp\!\Big[i\theta \sum_{m<j} n_m\Big],
\end{equation}
with $n_j = b_j^\dagger b_j = a_j^\dagger a_j$ being invariant under the transformation. Upon substituting Eq.\,\eqref{eq:jwtrans} in the bosonic canonical commutation relations, one can verify that anyonic operators satisfy the relations
\begin{flalign}
& a_j a_k - e^{i\theta\,\mathrm{sgn}(j-k)}\,a_k a_j = 0, \\
& a_j a_k^\dagger - e^{-i\theta\,\mathrm{sgn}(j-k)}\,a_k^\dagger a_j = \delta_{jk},
\end{flalign}
where $\mathrm{sgn}(x)$ denotes the sign function. Importantly, at $\theta = \pi$, these anyonic operators satisfy pseudo-fermion statistics: they behave like fermions when exchanging particles on different sites (i.e., acquire a $-1$ phase), but remain bosonic on-site, allowing double occupancy and thus, lacking Pauli exclusion. Because of the Jordan–Wigner string, the anyon operators are non-local in the bosonic basis: creating an anyon at site $j$ attaches a phase that depends on the occupation of all sites to its left \cite{2011_Keilmann}. Correspondingly, the kinetic energy part of the soft-core anyon-Hubbard model \eqref{eq:H_eff} can be rewritten as
\begin{equation}\label{eq:kinterm}
H_{\text{kin}} = -J \sum_j \big( a_j^\dagger a_{j-1} + \text{h.c.} \big),
\end{equation}
so that the density-dependent Peierls phases are entirely encoded in the algebra of the $a_j$ operators. Interactions that are powers of the density remain unchanged under the transformation.

\subsection{Gauge-potential view of the AHM}

The AHM can be seen as a theory of matter coupled to a dynamical gauge potential that transmutes the statistics of particles. To be more explicit, we shall recall that minimal coupling of Abelian gauge potentials to charged matter in the continuum is achieved via the change $\vec{p} \mapsto \vec{p} - q\vec{A} $ in the kinetic term, where $\vec{A}$ constitutes the spatial component of the gauge potential. In a tight-binding model, this prescription is replaced by the so-called Peierls substitution, which consists of the replacement $b^{\dagger}_{j}b_{k} \mapsto b^{\dagger}_{j} \,\mathcal{U}_{j,k}\, b_{k}$ in the hopping term, where $\mathcal{U}$ is a unitary operator known as the parallel transporter or link variable, which accounts for the presence of a non-null connection. The link variable might be written as $\mathcal{U}_{j,k} = \exp{(i \phi_{j,k})}$, where
\begin{equation}
    \phi_{j,k} = \frac{q}{\hbar} \int^{j}_{k} \vec{A}\cdot \text{d}\vec{x}\;,
\end{equation}
is the phase picked up by a particle when hopping between sites $k$ and $j$ in the presence of a gauge potential $\vec{A}$. When the gauge potential is a background field, $\phi_{j,k}$ reduces to a real-valued scalar known as the Peierls phase. However, when the gauge potential is dynamical, $\phi_{j,k}$ becomes operator-valued. In the AHM, we observe already from Eq. \eqref{eq:H_eff} that $\phi_{j,j-1} = -\theta\, n_{j}$ corresponds to the latter case, namely it is a density-dependent Peierls phase. 

The consequence of this is readily observable. The gauge-matter interaction can be written explicitly by expanding the link variable in series as
\begin{equation}
    \mathcal{U}_{j,j+1} = e^{i\theta n_{j}} = \sum_{q=0}^{\infty}\frac{(i\theta)^{q}}{q!} \sum_{m=0}^{q} S(q,m) (b^{\dagger}_{j})^{m}(b_{j})^{m} ,
\end{equation}
where $S(q,m)$ are the Stirling numbers of the second kind accounting for the normal ordering of bosonic operators. Using this expansion, and by virtue of Eq. \eqref{eq:jwtrans}, we can re-express the kinetic term in Eq. \eqref{eq:kinterm} as
\begin{equation}
H_{\text{kin}} = - J \sum_{j} \Big( b^{\dagger}_{j} b_{j-1} + e^{-i\theta} b^{\dagger}_{j} n_{j-1} b_{j-1}+ \text{h.o.t.}+ \text{h.c.}\Big)\;,
\end{equation}
where three-body and other higher-order terms (h.o.t.) can be neglected in the low-density regime, so that the second term in brackets can be considered the dominant statistical interaction as it is governed by $\theta$. This statistical interaction, which favors processes of density-assisted tunneling, will compete with the Hubbard repulsion and is intimately linked both with the presence of chiral binding and pseudo-fermionization. We recall that this new interaction is a many-body effect caused by the presence of non-trivial dynamical gauge potential disguised as a seemingly innocuous phase factor in the hopping term in Eq. \eqref{eq:H_eff}. The generalized Jordan-Wigner transformation reabsorbs this interaction in the commutation relations of anyons, something that might be referred to as statistical transmutation. It is worth noticing the similarity of this observation with the more familiar case of anyons in 2D, where the reabsorption of the statistical gauge potential in the commutation relations of matter fields is achieved by a singular gauge transformation, which plays the role of a higher dimensional analogue of the Jordan-Wigner transformation used here.

\section{Two-particle problem in the 1D AHM}

In the bosonic representation, the 1D AHM is described by the Hamiltonian
\begin{align}
    \hat{H}=-J\sum_{j}\left(\hat{b}_{j+1}^{\dagger} e^{-i \theta \hat{n}_{j+1}} \hat{b}_j + \text{h.c.}\right)
    + \frac{U}{2} \sum_j \hat{n}_j (\hat{n}_j-1),
    \label{eq:AHM}
\end{align}
with $\hat{b}_j$ ($\hat{b}^{\dagger}_j$) being a bosonic annihilation (creation) operator for a particle at site $j$, on-site interactions $U$, amplitude $J$ of the tunneling matrix elements, and number- or density-dependent Peierls phases characterized by the anyonic exchange angle $\theta$ \cite{2011_Keilmann}.

\subsection{Interaction strength}

The dilute regime in the 1D AHM can be understood from the low-energy two-body problem and is characterized by the two-body scattering length
\begin{equation}
    a_{\mathrm{1D}} = -\frac{1 + \cos(\theta)}{4[1 - \cos(\theta)] + 2U/J},
\end{equation}
which connects to the effective interaction strength $g_{\mathrm{1D}} = -2/(a_{\mathrm{1D}} m)$ describing a 1D Bose gas \cite{2015_Greschner, 2017_Zhang}. In particular, the effective interaction strength is always positive for $U \ge 0$ and diverges when $\theta \rightarrow \pi$, in line with the progressive pseudo-fermionization of 1D anyons.

\subsection{Two-particle configuration space}

\begin{figure}[tbh!]
    % //TODO possibly refer to Fig. 1 -> 2D conf. space diagram in main text
    \centering
    \includegraphics[width = \columnwidth]{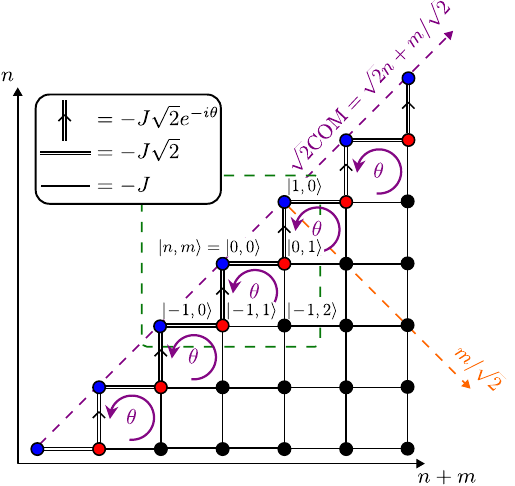}
    \caption{\textbf{Relative coordinate representation of the two-particle problem in the AHM.}
            The coordinates $n$ ($n+m$) represent the position of the leftmost (rightmost)
            particle such that $m$ denotes the relative distance between the two particles.
            The purple diagonal axis indicates the center-of-mass (COM) position $n+\frac{m}{2}$ 
            of the two particles while the relative distance $m$ is indicated by the
            orange axis perpendicular to the diagonal axis.
            Due to the indistinguishability of the particles, only the lower right triangle
            of the coordinate system is relevant. 
            The blue (red) circles indicate configurations of two-particle coincidences with $m=0$ (configurations where two neighboring sites are occupied with $m=1$).
            For the other black-colored sites, the particles are separated by at least one empty site ($m\geq2$).
            The different density-dependent tunneling matrix elements between these state configurations, as depicted by the bonds (see legend), are governed by the Hamiltonian 
            \eqref{eq:AHM-rel-coord-Ham}, which, in particular, leads to Aharonov-Bohm fluxes piercing the plaquettes along the diagonal, as indicated by the purple encircling arrows.
            The green dashed square contains all the configurations that are relevant 
            for the initial state of the experimental protocol of Fig.~3 in the main text.
            }
            \label{fig:relativecoord-AHM}
        \end{figure}

We study the two-particle problem in the 1D AHM~\eqref{eq:AHM} by mapping it to a single-particle problem in a two-dimensional configuration space~[\cref{fig:relativecoord-AHM}]. 
To this end, first note that any eigenstate of the system can be either expanded in terms of the bosonic Fock basis $\ket{\{n_j\}}$ with $\sum_j n_j=2$ or using the relative coordinate basis $\ket{n,m}$ for the configuration space, where $n$ and $n+m$ represent the position of the leftmost and the rightmost particle, respectively, with $m\ge0$ denoting their relative distance.
In the case of the AHM defined on an infinite 1D chain, we have $n \in \mathbb{Z}$ and $m\in\mathbb{N}_0$.
In the relative coordinate basis, the two-particle Hamiltonian reads
    \begin{align}
        \hat{H} = -&J \sum_{n} \Big[ \sqrt{2}\left\{
             e^{-i\theta}\ketbra{n,0}{n\!-\!1,1} +\ketbra{n,1}{n,0} 
            + \text{h.c.}  \right\} \nonumber \\
        + & \sum_{m\geq2}(\ketbra{n,m\!-\!1}{n,m} + \ketbra{n\!+\!1,m\!-\!1}{n,m}+ \text{h.c.})
        \Big]\nonumber \\
         + &U\sum_{n} \ketbra{n,0}{n,0},
    \label{eq:AHM-rel-coord-Ham}
    \end{align}
whose action on the two-particle configuration space is further depicted in~\cref{fig:relativecoord-AHM}. 
Thus, the two-particle problem described by the AHM Hamiltonian~\eqref{eq:AHM} can be mapped to a single-particle lattice model in two dimensions~\eqref{eq:AHM-rel-coord-Ham}. 
Note, in particular, that in this representation, the correlated density-dependent Peierls phases and on-site interactions become single-particle Peierls phases and on-site potential energies, respectively.
\subsubsection{Translational invariance and hybrid basis.} 
In the case of an underlying infinite 1D chain, the translational invariance of the chain translates itself to a translational invariance along the COM direction in the relative coordinate basis representation.
Thus, the COM quasi-momentum $q$ is conserved and becomes a good quantum number, such that we can block-diagonalize the Hamiltonian \eqref{eq:AHM-rel-coord-Ham} into sectors of fixed COM quasi-momentum $q$.
To this end, we first introduce hybrid basis states characterized by the sharp COM quasi-momentum $q$ and the relative position $m$
    \begin{align}
        \ket{q,m} = \sum_{n} e^{i q(n+\frac{m}{2})} \ket{n,m},
        \label{eq:AHM-hybrid-com-qmom-basis}
    \end{align}
where $n+\frac{m}{2}$ is the COM position of the two particles.
Using this, we may expand any eigenstate of~\eqref{eq:AHM-rel-coord-Ham} as
    \begin{align}
            \ket{ \Psi_{q,\kappa}} &= \sum_{m} c^{q,\kappa}_{m} \ket{q,m}=\sum_{n,m} c^{q,\kappa}_{n,m} \ket{n,m},
        \label{eq:AHM-rel-coord-expansion}
    \end{align}
with $c^{q,\kappa}_{n,m}=e^{i q(n+\frac{m}{2})}c^{q,\kappa}_{m}$,
where the quantum number $\kappa$ specifies the relative motion, distinguishing between energy eigenstates with the same COM quasi-momentum $q$. 
Now, representing the Hamiltonian \eqref{eq:AHM-rel-coord-Ham} in the hybrid basis~\eqref{eq:AHM-hybrid-com-qmom-basis} leads to
    \begin{align}
        \hat{H} = &\sum_{q} \hat{H}_q, \nonumber \\
             = &\sum_{q}\Big[-J\Big\{\sqrt{2}\big[
            2\cos{\left(\frac{q+\theta}{2}\right)}e^{-i\theta/2}\ketbra{q,0}{q,1} +\text{h.c.}\big] \nonumber \\
            &+ \sum_{m\geq2}2\cos(q/2)\big(\ketbra{q,m}{q,m\!-\!1}  +\text{h.c.}\big)\Big\}\nonumber \\
            &+U\ketbra{q,0}{q,0}\Big].
        \label{eq:AHM-rel-coord-Ham-q}
    \end{align}
As expected, states of different COM quasi-momentum decouple.
The Hamiltonian $\hat{H}_q$ can be interpreted as describing single-particle dynamics on a semi-infinite 1D chain with the finite edge located at $\ket{q,0}$, and the infinite direction extending towards increasing relative distances $m$ between the two particles $\ket{q,m\geq 0}$~[see \cref{fig:Effective-2body-spectrum}(b)].
This Hamiltonian is the starting point to solve for the energy $E_{q,\kappa}$ and the relative coordinate coefficients $c^{q,\kappa}_m$ of the two-particle wave function, which is governed by 
the time-independent Schrödinger equation $\hat{H} \ket{\Psi_{q,\kappa}} = E_{q,\kappa} \ket{\Psi_{q,\kappa}}$.

\subsubsection{Two-particle bound and scattering states.}
Using this representation \eqref{eq:AHM-rel-coord-Ham-q} and comparing the coefficients of the hybrid coordinate basis states $\ket{q,m}$, one can arrive at a linear set of recurrence relations for the relative coordinate coefficients $c^{q,\kappa}_m$ \cite{2024_Kwan}.
These can then be used to obtain the exact two-particle bound and scattering state solutions to the two-particle problem in the AHM, as it has been studied in previous works~\cite{2024_Kwan,2016_Cardarelli,2018_Greschner,2017_Zhang,2015_Greschner}.

The main features of these can be summarized as follows. 
Both the two-particle bound state and scattering state solutions are characterized by a plane-wave function along the COM direction, i.e., $c^{q,\kappa}_{n,m}=e^{i q (n+\frac{m}{2})}c^{q,\kappa}_{m}$, which is further modulated by a wave function $c^{q,\kappa}_{m}$ depending on the relative distance $m$ between the two particles.
For the two-particle bound states, the wave function $c^{q,\kappa}_{m}$ is exponentially decaying in the direction of increasing relative distance $m$ between the two particles, while for the two-particle scattering states, the wave function $c^{q,\kappa}_{m}$ is also spatially oscillatory and extended.

In sufficiently large systems the probability for two particles to meet is negligible, such that the two-particle wave function can be approximated by a product of two single-particle wave functions.
Hence, the two-particle energy spectrum for the scattering states can be obtained by the total energy of two free particles, $E_q(\kappa)=\epsilon(k_1)+\epsilon(k_2)=-4J\cos(q/2)\cos(\kappa)$, where $k_{1(2)}=q/2+(-)\kappa$ is the particle momentum of the first (second) particle, $\epsilon(k)=2J\cos(k)$ the free particle dispersion, $\kappa$ the relative momentum, and $q$ the COM quasi-momentum. 
This continuum of solutions is bounded from below and above,
$\abs{E_q(\kappa)} \leq 4J \cos\left(\frac{q}{2}\right)$, where the lower (upper) bound 
corresponds to the relative momenta $\kappa=0$ ($\kappa=\pm\pi$).

The two-particle energy spectrum for the bound states is obtained by solving the associated recurrence relations for the wave function $c^{q}_{m}$~\cite{2024_Kwan}, which yield the exact energy spectrum $E_q$ of the two-particle bound states for arbitrary $\theta$ and $U$ as a function of the COM quasi-momentum $q$ (see Refs.~\cite{2024_Kwan,2016_Cardarelli,2018_Greschner,2017_Zhang}).

The corresponding solution of the bound state energy spectrum for arbitrary statistical angles $\theta$ and on-site interaction $U$ is given by
\begin{align}
            E_{q,\pm}=&-\Bigg(U\left[\cos^2{\frac{q}{2}}-\cos^2\left(\frac{q+\theta}{2}\right) \right]
            \mp\cos^2\left(\frac{q+\theta}{2}\right) \nonumber\\ 
            &\times\sqrt{U^2+16\left[2\cos^2\left(\frac{q+\theta}{2}\right)-\cos^2{\frac{q}{2}}\right]}\ \Bigg)
            \nonumber\\
            &\times\left[2\cos^2{\left(\frac{q+\theta}{2}\right)}-\cos^2{\frac{q}{2}}\right]^{-1},
    \label{eq:AHM-bound-states-dispersion}
\end{align}
which for $U=0$ simplifies to
\begin{align}
    E_{q,\pm}=\pm\frac{
    4\cos^2{\left(\frac{q+\theta}{2}\right)}}
    {\sqrt{2\cos^2{\left(\frac{q+\theta}{2}\right)}-\cos^2{\frac{q}{2}}}}.
    \label{eq:AHM-bound-states-dispersion-Uzero}
\end{align}
Two-particle bound state solutions exist when the energy $E_q$ of these solutions is located outside the two-particle scattering state continuum, $\abs{E_q}\leq 4J\cos(q/2)$.
Note, in particular, that different from the conventional Bose-Hubbard model ($\theta=0$), two-body bound state solutions can also exist for $U=0$ in the AHM whenever $\theta\neq 0$, since the density-dependent Peierls phases mediate interactions between the bosons~\cite{2024_Kwan}.

\subsection{Chiral binding mechanism}\label{subsec:chiral-binding-mechanism}
\begin{figure}[tb!]
        \centering
        \includegraphics[width = \columnwidth]{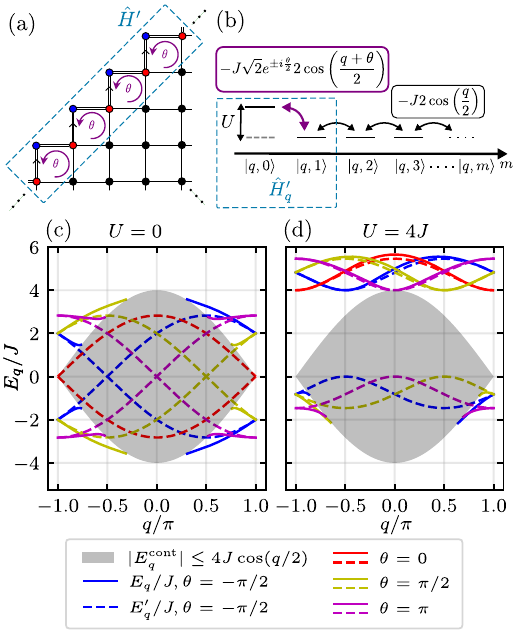}
        \caption{\textbf{Chiral binding mechanism and approximate two-particle energy spectrum.}
        (a) Illustration of the truncated two-particle configuration space (light-blue colored dashed rectangle with $m\leq 1$) to derive an approximate model $\hat{H}^{\prime }$ for the two-particle bound states in the AHM (cf.~\cref{fig:relativecoord-AHM}).
        (b) Illustration of the quasi-momentum diagonal two-particle AHM Hamiltonian $\hat{H}_q$~\eqref{eq:AHM-rel-coord-Ham-q} as acting on a semi-infinite one-dimensional chain spanned by the basis states $\ket{q,m}$. The truncated effective Hamiltonian $\hat{H}^{\prime }_q$ (light-blue colored dashed box) corresponding to the approximation of (a) contains only the configurations $\ket{q,0}$ and $\ket{q,1}$.
        For $U=0$ (c) and $U=4J$ (d), we show the approximate two-particle bound state spectrum $E^{\prime}_q$ \eqref{eq:AHM-rel-eff-energy-ideal} (dashed) together with the exact two-particle bound state energy spectrum $E_q$ \eqref{eq:AHM-bound-states-dispersion} (solid) for different statistical angles $\theta$ (colors).
        The scattering-state continuum $\abs{E^{\text{cont}}_q} \leq 4J\cos(q/2)$ is shown as the gray shaded area.}
        \label{fig:Effective-2body-spectrum}
    \end{figure}
For $U=0$, we provide an intuitive explanation of the mechanism giving rise to two-particle bound states. 
In the limit of large systems, we find delocalized solutions in the bulk of the 2D lattice of \cref{fig:relativecoord-AHM}, which correspond to the scattering state solutions.
Their kinetic energy is minimal for zero quasi-momentum and reads $-4J$ 
[that is $-2J$ for the delocalization in each of the two directions/particles].

We can now also make a simple ansatz for potential bound states, 
by restricting the wave function to the red and blue-colored sites~[\cref{fig:Effective-2body-spectrum}(a)], defining a one-dimensional system, i.e., the particles are only allowed to be separated by at most one site ($m \leq 1$).
This corresponds to the motion along a simple one-dimensional tight-binding chain where both particles stay together and move in one direction by a two-step process:
(i) every other tunneling process is associated with a finite Peierls phase $\pm\theta$, and (ii) the tunneling matrix elements are enhanced bosonically by a factor of $\sqrt{2}$. 
The effective Hamiltonian, denoted as $H^{\prime}$, corresponding to this approximation reads as~\eqref{eq:AHM-rel-coord-Ham} but truncated to $m \leq 1$~[\cref{fig:Effective-2body-spectrum}(a)].

In the case of an infinite 1D chain, this truncated effective Hamiltonian becomes translationally invariant with respect to the two-sublattice site unit cell of the system, such that it can be solved analytically by representing the effective Hamiltonian in the hybrid quasi-momentum space basis~\eqref{eq:AHM-hybrid-com-qmom-basis} resulting in $\hat{H}^{\prime }_q$.
The effective Hamiltonian $\hat{H}^{\prime }_q$ reads as~\eqref{eq:AHM-rel-coord-Ham-q} but truncated to processes with $\ket{q,m\leq 1}$, describing a two-level system at each COM quasi-momentum $q$~[\cref{fig:Effective-2body-spectrum}(b)], i.e.,
$\hat{H}^{\prime } = \sum_{q} \hat{H}^{\prime }_q$ with
$\hat{H}^{\prime }_q=h_0(q)\mathbbm{1}_{2}+\bm{h}(q)\cdot \hat{\bm{\sigma}}$ with $\hat{\bm{\sigma}}=(\sigma_x,\sigma_y,\sigma_z)^T$ denoting the vector of Pauli matrices, $\mathbbm{1}_2$ the $2\times 2$ identity matrix, and $\bm{h}(q)=(h_x(q),h_y(q),h_z(q))^T$ a vector on the Bloch sphere with the north and south poles corresponding to the two states $\ket{q,0}$ and $\ket{q,1}$. Here, $h_0(q)=h_z(q)=U/2$ and $(h_x(q),h_y(q)) = -J2\sqrt{2}\cos{\Big(\frac{q+\theta}{2}\Big)}(\cos(\theta/2),\sin(\theta/2))$.
Thus, the approximate two-particle bound-state energy spectrum $E^{\prime}_{q,\pm} =h_0(q)\pm \abs{\bm{h}(q)}$ is given by
\begin{align}
    E^{\prime}_{q,\pm}/J = \frac{U+\pm\sqrt{16+U^2+16\cos(q+\theta)}}{2},
    \label{eq:AHM-rel-eff-energy-ideal}
\end{align}
which for $U=0$, further simplifies to 
    \begin{align}
    E^{\prime}_{q,\pm}/J
    % =\pm 2\sqrt{1+\cos(q+\theta)}
    =\pm2\sqrt{2}\cos{\left(\frac{q+\theta}{2}\right)}.
        \label{eq:AHM-rel-eff-energy-ideal-Uzero}
    \end{align}
In \cref{fig:Effective-2body-spectrum}(c)-(d), for $U=0$ and $U=4J$ and
for various statistical angles $\theta$, we compare the exact~\eqref{eq:AHM-bound-states-dispersion} and approximate~\eqref{eq:AHM-rel-eff-energy-ideal} two-particle bound-state spectrum together, where we find good agreement as the approximate band structure lies outside the scattering-state continuum but above (below) the exact solutions of the lower (upper) bound-state branches.
Note that a similar approximate two-particle spectrum has been derived in~\cite{2016_Cardarelli} for deeply bound pairs of spinful fermions with engineered interactions.

For $U=0$, here, we see that, as a result of the Peierls phases (i), the bound states minimize their kinetic energy for the finite COM quasi-momentum $q=-\theta$~\eqref{eq:AHM-rel-eff-energy-ideal-Uzero}.
In turn, the scattering states hardly feel this Peierls phase at the edge,
so that at finite $q$ their energy is increased relative to their ground state at $q=0$. 
The scattering states still have the advantage that they can lower their kinetic energy by delocalizing also with respect to their relative distance with zero quasi-momentum in this direction.
However, this advantage is partially compensated by the bosonic enhancement (ii), 
which allows the co-moving bound particles to lower their energy by a factor of $\sqrt{2}$ 
to $2\sqrt{2}$~\eqref{eq:AHM-rel-eff-energy-ideal-Uzero} for $U=0$.

As a consequence of (i) and (ii), for $U=0$ the kinetic energy of the bound state takes 
a minimal value of $-2\sqrt{2}J$  \eqref{eq:AHM-rel-eff-energy-ideal},
which can lie, for $\theta\neq 0$ and finite $q$, outside the energy of the scattering 
continuum, as shown in \cref{fig:Effective-2body-spectrum}.
From this approximate estimate, for sufficiently large $\theta$ and sufficiently large quasi-momenta $q$, we expect stable bound-state solutions below and above the scattering continuum.
Since the binding mechanism requires finite quasi-momenta, it is a non-equilibrium effect protected by momentum conservation, i.e., bound states can only be probed when starting from populated excited states of the system away from the global ground state which is always given by a (non-bound) scattering state at $q=0$.

We conclude with a few comments on our approximation. 
First, we note that the energy $E_{q,\pm}$ of the system \eqref{eq:AHM-rel-eff-energy-ideal} 
is independent of the statistical angle $\theta$, i.e., $\theta$ only determines the
location of the energy minimum of the approximate band structure.
This is expected as there are no Fock space loops in configuration space enclosing a non-trivial flux in our approximation of $\hat{H}^{\prime }$~[\cref{fig:Effective-2body-spectrum}(a)].
Second, we note that, for $q=\pm \pi$, our approximation and hence the two-particle energy spectrum becomes the exact, which is a direct consequence of a destructive interference effect dissecting the configuration space into two disconnected parts, the edge ($m\leq 1$) and the bulk ($m\geq 2$) of the configuration space~[\cref{fig:Effective-2body-spectrum}(a)-(b)].  
This is easily seen by comparing $\hat{H}_q$~\eqref{eq:AHM-rel-coord-Ham-q} and $E_{q,\pm}~$\eqref{eq:AHM-bound-states-dispersion} to their truncated approximative versions $\hat{H}^{\prime }_q$ and $E_{q,\pm}^{\prime }$~\eqref{eq:AHM-rel-eff-energy-ideal}, respectively, which coincide
at $q=\pm \pi$ (since the tunneling matrix element connecting the edge to the bulk vanishes identically, i.e., $\cos(q/2)\rvert_{q=\pm \pi}=0$~[\cref{fig:Effective-2body-spectrum}]).
Third, we can directly deduce from $\hat{H}^{\prime }_q$ that, for $U=0$, the two-particle solutions consist of an equal weight superposition of the two configurations $\ket{q,0}$ and $\ket{q,1}$, as $\bm{h}(q)$ only points along the equatorial $x$-$y$-plane on the Bloch sphere, whereas for $U\neq 0$ there is a stronger population of the $\ket{q,0}$ ($\ket{q,1}$) configuration for $U>0$ ($U<0$) since the vector $\bm{h}(q)$ points into the upper (lower) hemisphere of the Bloch sphere.

\subsection{Observables at short times}
\begin{figure}[t!]
    \centering
    \hspace*{-.5cm}  % Adjust this value as needed to shift the figure to the left
    \includegraphics[width=\columnwidth]{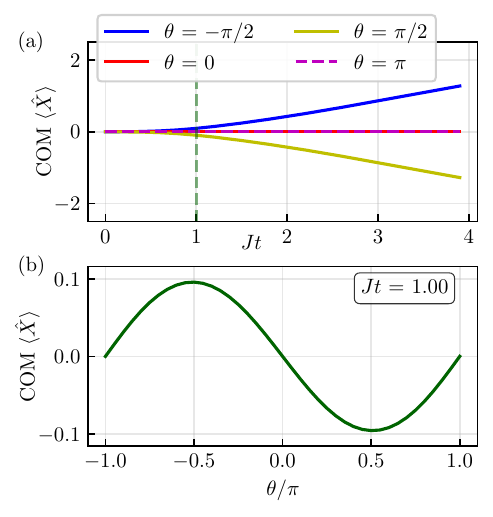}
    \caption{\textbf{Asymmetric two-particle expansion dynamics.}
        (a) Center-of-mass (COM) motion $\ep{\hat{X}}$~\eqref{eq:COM-op} over the evolution time
         for various statistical angles $\theta \in \{-\pi/2,\,0,\,\pi/2,\,\pi\}$ (colors). 
         (b) COM drift $\ep{\hat{X}}$ at the fixed evolution time $Jt=1$ (dashed vertical line in (a)) as a function of the statistical angle $\theta$.
         In both (a) and (b), the initial state is given by the
         ground state $\ket{\Phi}$ of $\hat{H}$~\eqref{eq:AHM} for $N=2$ bosons on $L'=3$ lattice sites embedded in a larger lattice of $L=31$ sites, which is evolved under the Hamiltonian $\hat{H}$~\eqref{eq:AHM} for $U=0$.}
    \label{fig:com_motion_theta_t}
\end{figure}

In the experiment, the protocol for the asymmetric expansion dynamics starts from the adiabatically prepared two-particle ($N=2$) ground state confined to three lattice sites ($L'=3$) with zero on-site interaction ($U=0$).
Then, the confinement is quenched to zero, and the system is let to freely expand under the AHM lattice of larger size $L\gg L'$.
A numerical simulation of this protocol is performed via exact diagonalization and depicted in \cref{fig:com_motion_theta_t}.
The resulting COM drifts $\ep{\hat{X}}$~\eqref{eq:COM-op} are depicted over time for various statistical angles $\theta$ in \cref{fig:com_motion_theta_t}(a).
In \cref{fig:com_motion_theta_t}(b), we show COM drifts over the statistical angle $\theta$
at a fixed evolution time.
In contrast to previous experiments~\cite{2024_Kwan}, starting from a two-particle product state $\ket{\ldots0110\ldots}$, here a finite asymmetric COM drift already appears for $U=0$. 
The reason for this asymmetric drift will be explained in the following as well as its dependence 
on the statistical angle $\theta$. 

\subsubsection{Two-particle ground state on three sites}
\begin{figure*}[t!]
    \centering
    \includegraphics[width=2\columnwidth]{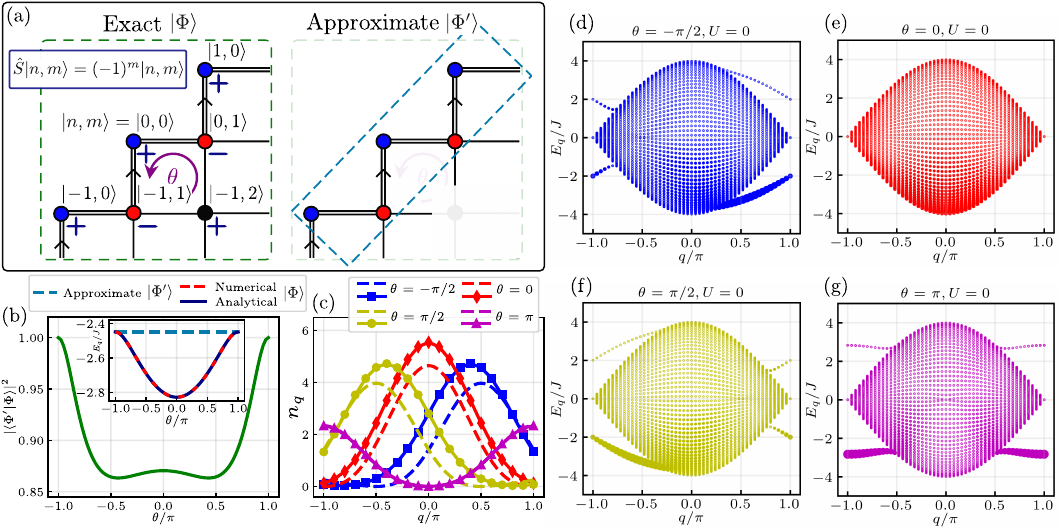}
    \caption{\textbf{Two-particle AHM ground state on three sites for $U=0$.}
    (a) (Left) The 6-site configuration space representation~(cf.~\cref{fig:relativecoord-AHM}) relevant for the initial two-particle ground state. The chirality $(-1)^m$~\eqref{eq:chiral-symmetry-op-rel-coord} of each basis state is indicated by the $\pm$ signs, which is used to compute the exact ground state $\ket{\Phi}$~\eqref{eq:AHM-ground-state-chiral-basis}.
    (Right) Illustration of the finite 5-site chain (light-blue colored dashed box) in the configuration space used to obtain the approximate ground state $\ket{\Phi^{\prime}}\equiv \ket{\Phi^{\prime}_{1}}$~\eqref{eq:two-part-approx-5-site-chain-eigenstates}.
    (b) Overlap $\abs{\braket{\Phi^{\prime}}{\Phi}}^2$ of the exact ground state $\ket{\Phi}$ with the approximate ground state $\ket{\Phi^{\prime}}$ as a function of the statistical angle $\theta$. 
    The inset shows the energy of the approximate ground state $E_{q_1}=-\sqrt{6}J$ and of the exact ground state as a function of $\theta$. 
    The dashed red line shows the numerical value obtained by exact diagonalization (ED), and the solid blue line shows the analytical value $E_{1,-}/J = -\sqrt{5+\sqrt{5+4\cos(\theta)}}$~\eqref{eq:AHM-energy-eigenvalues-3sites}.
    (c) COM quasi-momentum distribution $n_q= \abs{\sum_m\braket{q,m}{\Phi}}^2$ of
    the exact (solid line with markers) and approximate (dashed line) ground state for various statistical angles $\theta$ (colors).
    (d)-(g) Overlap $\mathcal{O}=\abs{\braket{\Psi^q}{\Phi}}^2$ of the exact ground state with the entire two-particle spectrum $\{\ket{\Psi^q}\}$~\eqref{eq:AHM-rel-coord-Ham-q} of the AHM for various statistical angles $\theta=-\pi/2$ (c), $\theta=0$ (d), $\theta=\pi/2$ (e), and $\theta=\pi$ (f).
    The magnitude of the overlap is illustrated by the size of the area of the filled circles (scales quadratically).}
    \label{fig:ovlap-exact-gs-and-approx-energy-mom-dist}
\end{figure*}

{\label{sec:exact-ground-state}}
In the experiment, the initial state is prepared in the ground state 
$\ket{\Phi}$ of two bosons on three lattice sites for the AHM \eqref{eq:AHM} for $U=0$.
Using the relative coordinate basis representation, the correct ground state 
can be represented as a superposition of six basis states,
    \begin{align}
        \ket{\Phi} = \sum_{n=-1}^1d_{n0}\ket{n,0}+\sum_{n=-1}^{0}d_{n1}\ket{n,1}+d_{-12}\ket{-1,2},
        \label{eq:AHM-correct-gs-superpos-new}
    \end{align}
which are contained within by the green dashed square [left panel in~\cref{fig:ovlap-exact-gs-and-approx-energy-mom-dist}(a)] with 
$d_{nm}=\braket{n,m}{\Phi}$ being the expansion coefficients of the ground state in the relative coordinate basis.
In the Fock state representation $\ket{n_1\,n_2\,n_3}$, these six basis states correspond to 
$\ket{200} = \ket{-1,0}$, $\ket{020} = \ket{0,0}$, $\ket{002} = \ket{1,0}$, 
$\ket{110} = \ket{-1,1}$, $\ket{011} = \ket{0,1}$, $\ket{101} = \ket{-1,2}$.
In the following, we will first briefly derive a simple approximate expression for this ground state, which connects to the intuitive picture of the chiral binding mechanism presented above.
Afterward, we will then present the derivation of the exact analytical expression for this ground state, which is based on the chiral symmetry of the AHM Hamiltonian~\cite{2025_Theel}.

\emph{Approximate two-particle ground state on three sites.}
A simple approximation to the two-particle ground state on three sites can be obtained by neglecting the configuration $\ket{-1,2}$ in~\eqref{eq:AHM-correct-gs-superpos-new}, by which the system reduces to a simple single-particle problem on a one-dimensional 5-site chain, as illustrated in the right panel of~\cref{fig:ovlap-exact-gs-and-approx-energy-mom-dist}(a).
After removing the remaining Peierls phases by suitable gauge transformations (which is allowed since the non-trivial flux is no longer present in the approximation), the Hamiltonian describing this approximate system is simply given by $\hat{H}= -J\sqrt{2}\sum_{\ell=1}^{M-1}(\ketbra{\ell+1}{\ell}+\text{h.c.})$. Here, we have labeled the configurations of the $5$-site chain ($M=5$) by $\ell \in \{1,2,3,4,5\}$ with $\ell$ increasing from the lower left blue-colored configuration upwards, i.e., $\ket{\ell=1}=\ket{-1,0}$, $\ket{\ell=2}=\ket{-1,1}$, etc.~[\cref{fig:ovlap-exact-gs-and-approx-energy-mom-dist}(a)].
The eigenstates of such a tight-binding chain with open boundary conditions are well-known and given by the superposition of two plane waves corresponding to the same energy $E_q=-2\sqrt{2}J\cos(q/2)=E_{-q}$ of quasimomentum $q$ and $-q$, with the boundary conditions selecting the allowed quasi-momenta $q=q_s$ for $s\in\{1,\ldots,M\}$, which are given by
    \begin{align}
        \ket{\Phi^{\prime}_s}&=\sqrt{\frac{2}{M+1}}\sum_{\ell=1}^M e^{-i \frac{\theta}{2} (\ell+(\ell \bmod 2))}\sin\left(\frac{q_s \ell}{2}\right)\ket{\ell},\nonumber \\
        &\text{with} \ 
        q_s=\frac{2s\pi }{M+1},\ \text{for} \ s\in\{1,\ldots,M\},
        \label{eq:two-part-approx-5-site-chain-eigenstates}
    \end{align}
where we have reintroduced the Peierls phases to work in the original gauge of the model~\eqref{eq:AHM}. 
The energy eigenvalues corresponding to the approximate eigenstates~\eqref{eq:two-part-approx-5-site-chain-eigenstates} read $E_{q_s}=-2\sqrt{2}J\cos(q_s/2)$.
The approximate two-particle ground state is then given by $\ket{\Phi^{\prime}}\equiv \ket{\Phi^{\prime}_{s=1}}$~\eqref{eq:two-part-approx-5-site-chain-eigenstates} corresponding to
the lowest energy eigenvalue $E_{q_1}=-2\sqrt{2}J\cos(\pi/6)=-\sqrt{6}J$.

We compare this approximate ground state $\ket{\Phi^{\prime}}$ to the exact ground state $\ket{\Phi}$ in~\cref{fig:ovlap-exact-gs-and-approx-energy-mom-dist}(b)-(c).
In \cref{fig:ovlap-exact-gs-and-approx-energy-mom-dist}(b), we show the overlap $\abs{\braket{\Phi^{\prime}}{\Phi}}$ between the approximate and exact ground state over the statistical angle $\theta$, where the inset further shows the energy of the exact and approximate ground state as a function of $\theta$.
We find a significant overlap for all values of $\theta$, in particular for $\theta=\pm \pi$, where the overlap becomes unity, implying that the approximate ground state $\ket{\Phi^{\prime}}$ becomes exact. This is due to the inversion symmetry of the AHM and a destructive interference effect at $\theta=\pm \pi$, leading to no population of the $\ket{-1,2}$ configuration in the exact ground state $\ket{\Phi}$.
Next, by projecting onto the set of all possible
hybrid COM quasi-momentum states $\ket{q,m}$~\eqref{eq:AHM-hybrid-com-qmom-basis}, 
the COM quasi-momentum distribution $n_q= \abs{\sum_m\braket{q,m}{\Phi^{\prime}}}^2$ of the approximate ground state is computed and depicted in~\cref{fig:ovlap-exact-gs-and-approx-energy-mom-dist}(c) for various angles $\theta$.
We find that $\ket{\Phi^{\prime}}$ is centered around a finite COM quasi-momentum $q=-\theta$ around which the two-body bound states are also found, matching the intuitive picture of the chiral binding mechanism presented above~[\cref{fig:Effective-2body-spectrum}].

\emph{Exact two-particle ground state on three sites.}
We can derive an exact analytical expression for the two-particle ground state of the AHM by utilizing the chiral symmetry of the AHM Hamiltonian~\eqref{eq:AHM} without on-site interaction ($U=0$), which has recently been revealed in Ref.~\cite{2025_Theel}. 

In any particle-number-conserving bosonic system, as in the AHM \eqref{eq:AHM}, the chiral symmetry corresponds to a unitary operator $\hat{S}$ that is an involution, i.e., $\hat{S}^2=\mathbbm{1}$, and anticommutes with the Hamiltonian, i.e., $\{\hat{H},\hat{S}\}=0$~\cite{2025_Theel,2015_Ludwig,2017_Yu,2024_Xiao}.
The involution of the chiral symmetry, $\hat{S}^2=\mathbbm{1}$, leads to a decomposition of the finite-dimensional Hilbert space $\mathcal{H}$ into \emph{chiral} eigenspaces $\mathcal{H}_{\pm}$, i.e., $\mathcal{H}=\mathcal{H}_+ \oplus \mathcal{H}_-$. 
These chiral eigenspaces $\mathcal{H}_{\pm}$, which are spanned by the eigenstates of $\hat{S}$ with eigenvalues $\pm 1$, have dimension $d_{\pm}=\dim \mathcal{H}_{\pm}$.
It follows from the anticommutation relations that $\hat{H}$ is antiblock-diagonal in the chiral basis and that the spectrum of $\hat{H}$ is symmetric around zero energy,
where, moreover, the zero-energy subspace $\mathcal{H}_0$ of $\hat{H}$ has a dimension $d_0=\dim \mathcal{H}_0 \geq |d_+-d_-|$~\cite{1986_Sutherland,1989_Lieb,2017_Ramachandran,2025_Nicolau}.
Here, for the AHM~\eqref{eq:AHM}, the chiral symmetry operator $\hat{S}$ is given by $\hat{S} = \exp\left(i\pi \sum_{j=1}^L j\hat{n}_{j}\right)$~\cite{2025_Theel},
with $\hat{n}_j=\hat{b}^{\dagger}_j\hat{b}_j$ being the bosonic number operator on site $j$ and $L$ the total number of lattice sites on the 1D chain.
It transforms the bosonic operators as $\hat{S}\hat{b}_j\hat{S}=(-1)^j \hat{b}_j$, from which we see that $\hat{S}\hat{H}\hat{S}=-\hat{H}$ holds for the AHM Hamiltonian \eqref{eq:AHM} with $U=0$.
In the configuration space, the chiral symmetry operator assigns opposite chirality of $\pm 1$ onto adjacent configurations, which differ by a single tunneling event.

In the two-particle configuration space representation with the basis states $\ket{n,m}$, the action of the chiral symmetry operator $\hat{S}$ is given by
\begin{align}
    \hat{S}\ket{n,m} = e^{i\pi (2n+m)}\ket{n,m}=e^{i\pi m}\ket{n,m}.
    \label{eq:chiral-symmetry-op-rel-coord}
\end{align}
Here, in the six-dimensional Hilbert space of the ground state $\ket{\Phi}$~\eqref{eq:AHM-correct-gs-superpos-new}, we have four (two) basis states with positive (negative) chirality, i.e., $\mathcal{H}_+=\text{span}\{\ket{-1,0},\ket{-1,2},\ket{0,0},\ket{1,0},\}$, and $\mathcal{H}_-=\text{span}\{\ket{-1,1},\ket{0,1}\}$, respectively,
such that $d_+=4$ and $d_-=2$ as illustrated~\cref{fig:ovlap-exact-gs-and-approx-energy-mom-dist}(a).
This guarantees the existence of a zero-energy subspace $\mathcal{H}_0$, which is at least two-dimensional, $d_0 \geq |d_+-d_-|=2$. 
The chiral symmetry operator $\hat{S}$ in its eigenbasis thus reads
\begin{align}
    \hat{S} = \begin{pmatrix}
        \mathbbm{1}_{4\times 4} & 0 \\
        0 & -\mathbbm{1}_{2\times 2}
    \end{pmatrix},
    \label{eq:chiral-symmetry-matrix}
\end{align}
where $\mathbbm{1}_{n\times n}$ is the $n\times n$ identity matrix and where the upper left (lower right) block corresponds to the positive (negative) chirality subspace $\mathcal{H}_+$ ($\mathcal{H}_-$).

The AHM Hamiltonian~\eqref{eq:AHM} for the two-particle problem on three sites with $U=0$, in the chiral eigenbasis 
% ($\{\ket{-1,0},\ket{-1,2},\ket{0,0},\ket{1,0},\ket{-1,1},\ket{0,1}\}$) 
is given by \begin{align}
    \hat{H} &= \begin{pmatrix}
        0 & A \\
        A^{\dagger} & 0
    \end{pmatrix}, \ \text{with} \ 
    A= -J\begin{pmatrix}
        \sqrt{2} & 0 \\
        1 & 1 \\
        \sqrt{2}e^{-i\theta} & \sqrt{2} \\
        0 & \sqrt{2}e^{-i\theta}
    \end{pmatrix}.
    \label{eq:AHM-Hamiltonian-matrix-3sites-chiral-basis}
\end{align}
As expected, the Hamiltonian matrix $\hat{H}$ admits an anti-block-diagonal in accordance with the chiral symmetry, where the $4\times 2$ matrix $A\in \mathbb{C}^{4\times 2}$ describes the coupling between the negative and positive chirality subspaces, i.e., $A: \mathcal{H}_- \to \mathcal{H}_+$.
To obtain the energy eigenvalues of the system, we can square the Hamiltonian matrix~\eqref{eq:AHM-Hamiltonian-matrix-3sites-chiral-basis}, which leads to a block-diagonal form
\begin{align}
    \hat{H}^2 = \begin{pmatrix}
        AA^{\dagger} & 0 \\
        0 & A^{\dagger}A
    \end{pmatrix}.
    \label{eq:AHM-Hamiltonian-matrix-3sites-chiral-basis-squared}
\end{align}
The eigenvalues of $\hat{H}^2$ are given by the eigenvalues of the two blocks $AA^{\dagger}\in\mathbb{C}^{4\times 4}$ and $A^{\dagger}A\in \mathbb{C}^{2\times 2}$.
Taking the square root of these eigenvalues then yields the exact energy eigenvalues of the system in the chiral basis (or any other unitarily equivalent basis).
Moreover, the \emph{non-zero} eigenvalues of $AA^{\dagger}$ and $A^{\dagger}A$ coincide and are given by the squares of the non-zero singular values of $A$~\cite{2012_Horn}.
This stems from the fact that we can perform a singular value decomposition (SVD) of $A$, as
\begin{align}
    A=U\Sigma V^{\dagger}=\sum_{i=1}^{r}\sigma_i \ketbra{u_i}{v_i},
    \label{eq:AHM-matrix-A-SVD}
\end{align}
where $r=\text{rank}(A)$ is the rank of $A$, $\Sigma\in \mathbb{R}^{4\times 2}$ is the rectangular diagonal matrix containing the singular values $\sigma_i$ of $A$, and $U\in \mathbb{C}^{4\times 4}$ and $V\in \mathbb{C}^{2\times 2}$ are unitary matrices containing the left and right singular vectors $\ket{u_i}$ and $\ket{v_i}$ of $A$, respectively.
Note that here $r=2$, such that the zero-energy subspace is exactly two-dimensional $d_{0}=d_{+} +d_{-} - 2r=2$, resulting in $H_c$ having $d-d_0=4$ non-zero energy eigenvalues. 
To obtain the non-zero eigenvalues of $\hat{H}$, we first compute $A^{\dagger}A$, which reads
\begin{align}
    A^{\dagger}A = J^2\begin{pmatrix}
        5 & c \\
        c^* & 5
    \end{pmatrix},\quad c=1+2e^{i\theta},
    \label{eq:AHM-matrix-AAdagger}
\end{align}
where $c^*$ denotes its complex conjugate of the complex number $c\equiv c(\theta)$, which depends on the statistical angle $\theta$. 
The eigenvalues of $A^{\dagger}A$ are given by $\lambda_{1,2}/J^2=(5\pm |c|)=5\pm \sqrt{5+4\cos(\theta)}$, which directly yields the four non-zero energy eigenvalues of the Hamiltonian $\hat{H}$ as
\begin{align}
    E_{1,\pm}/J &= \pm\sqrt{5+\sqrt{5+4\cos(\theta)}},\nonumber \\
    E_{2,\pm}/J &= \pm\sqrt{5-\sqrt{5+4\cos(\theta)}},
    \label{eq:AHM-energy-eigenvalues-3sites}
\end{align}
where the remaining two eigenvalues of $\hat{H}$ are zero, corresponding to the two-dimensional ($d_0=2$) zero-energy subspace $\mathcal{H}_0$. 
In~\cref{fig:ovlap-exact-gs-and-approx-energy-mom-dist}(b) we show the exact ground state energy $E_{1,-}/J=-\sqrt{5+\sqrt{5+4\cos(\theta)}}$~\eqref{eq:AHM-energy-eigenvalues-3sites} as a function of $\theta$, where a comparison to the exact diagonalization (ED) results shows perfect agreement.

To obtain the corresponding eigenstates of $\hat{H}$, we proceed as follows.
First, we compute the eigenvectors of $A^{\dagger}A$, corresponding to the eigenvalues $\lambda_{1,2}$, which are given by
\begin{align}
    \ket{v_1} = \frac{1}{\sqrt{2}}\begin{pmatrix}
        e^{i\zeta} \\ 1
    \end{pmatrix},\ \ket{v_2} = \frac{1}{\sqrt{2}}\begin{pmatrix}
        e^{i\zeta} \\ -1
    \end{pmatrix},  
    \label{eq:AHM-eigenvectors-AAdagger}
    \\
    \text{with} \ e^{i\zeta}=\frac{c}{\abs{c}}=\frac{1+2e^{i \theta}}{\sqrt{5+4\cos(\theta)}},
    \nonumber
\end{align}
where for brevity we have introduced the phase $e^{i\zeta}$ of the complex number $c$, where in the following it is understood that $\zeta\equiv\zeta(\theta)$ depends on $\theta$.
Now, from the SVD of $A$~\eqref{eq:AHM-matrix-A-SVD}, we know that $A^\dagger A=V\Sigma^\dagger \Sigma V^\dagger$, such that the eigenvectors $\ket{v_i}$ of $A^{\dagger}A$ are also the right singular vectors of $A$, 
and we also know that $A\ket{v_i}=\sigma_i \ket{u_i}$, from which we can directly obtain the left singular vectors of $A$ as $\ket{u_i}=(A\ket{v_i})/\sigma_i$, with $\sigma_i>0$.
Here, the two left singular vectors $\ket{u_1}$ and $\ket{u_2}$ of $A$, using $\sigma_i\equiv E_{i,+}$~\eqref{eq:AHM-energy-eigenvalues-3sites}, can be computed as
\begin{align}
    \ket{u_1} &= \frac{A\ket{v_1}}{E_{1,+}}=\frac{-J}{ E_{1,+}\sqrt{2}}
    \begin{pmatrix}
        \sqrt{2}e^{i\zeta} \\ e^{i\zeta}+1 \\ \sqrt{2}(e^{i(\zeta-\theta)}+1) \\ 
        \sqrt{2} e^{-i\theta}
    \end{pmatrix},\nonumber \\
    \ket{u_2} &=\frac{A\ket{v_2}}{E_{2,+}}= \frac{-J}{E_{2,+}\sqrt{2}}
    \begin{pmatrix}
        \sqrt{2}e^{i\zeta} \\ e^{i\zeta}-1 \\ \sqrt{2}(e^{i(\zeta-\theta)}-1) \\ 
        -\sqrt{2} e^{-i\theta}
    \end{pmatrix}.
    \label{eq:AHM-left-singular-vectors-A}
\end{align}
By this, we have obtained the analytical singular value decomposition of $A= E_{1,+} \ketbra{u_1}{v_1}+E_{2,+} \ketbra{u_2}{v_2}$. 
This allows us now to construct the four energy eigenstates of $\hat{H}$, 
corresponding to the four non-zero energy eigenvalues $E_{i,\pm}$~\eqref{eq:AHM-energy-eigenvalues-3sites}, which can be shown to be always given in the form of~$\ket{\Psi_{i,\pm}} = \frac{1}{\sqrt{2}}\begin{pmatrix}
    \ket{u_i} \\ \pm \ket{v_i}
    \end{pmatrix}$,
where $\ket{u_i}$ and $\ket{v_i}$ are the left and right singular vectors from the SVD of $A$~\eqref{eq:AHM-matrix-A-SVD}.

Here, we are only interested in the ground state $\ket{\Phi}\equiv \ket{\Psi_{1,-}}$ of the system, corresponding to the lowest energy eigenvalue $E_{1,-}$~\eqref{eq:AHM-energy-eigenvalues-3sites}, which, in the chiral basis, reads
\begin{align}
    \ket{\Phi} = \frac{1}{\sqrt{2}}\begin{pmatrix}
        \ket{u_1} \\ -\ket{v_1}
    \end{pmatrix} = 
    \frac{-J}{2E_{1,+}}\begin{pmatrix}
        \sqrt{2}e^{i\zeta} \\ (e^{i\zeta}+1) \\ \sqrt{2}(e^{i(\zeta-\theta)}+1) \\ 
        \sqrt{2} e^{-i\theta} \\ E_{1,+}e^{i\zeta}/J \\ E_{1,+}/J
    \end{pmatrix}.
    \label{eq:AHM-ground-state-chiral-basis}
\end{align}
This ground state solves the following eigenvalue equation $\hat{H}\ket{\Phi}=E_{1,-}\ket{\Phi}$ in the chiral basis.
Note that we retrieve that the exact ground state~\eqref{eq:AHM-ground-state-chiral-basis} has no population on the $\ket{-1,2}$ configuration, i.e., $d_{-1,2}=\braket{-1,2}{\Phi}=\frac{-J}{2E_{1,+}}(e^{i\zeta(\theta)}-1)=0$ for $\theta=\pm \pi$, corroborating that the approximate ground state $\ket{\Phi^{\prime}}$~\eqref{eq:two-part-approx-5-site-chain-eigenstates} becomes exact in this case~[\cref{fig:ovlap-exact-gs-and-approx-energy-mom-dist}(b)].

As for the approximate ground state $\ket{\Phi^{\prime}}$ above, we compute the COM quasi-momentum distribution of this ground state $n_q= \abs{\sum_m\braket{q,m}{\Phi}}^2$, which is also shown in~\cref{fig:ovlap-exact-gs-and-approx-energy-mom-dist}(c) for various statistical angles $\theta$.
We find that the exact ground state $\ket{\Phi}$ is centered around a finite COM quasi-momentum, which is slightly shifted from $q=-\theta$ of the approximate ground state $\ket{\Phi^{\prime}}$ for $\theta\neq 0,\pm \pi$. 
Being centered around a finite $q$ implies that the ground state exhibits a large overlap with the two-body bound states at the same COM quasi-momentum.

This is further corroborated by computing the overlap $\mathcal{O}=\abs{\braket{\Psi^q}{\Phi}}^2$
of the exact ground state with the entire two-particle spectrum $\{\ket{\Psi^q}\}$~\eqref{eq:AHM-rel-coord-Ham-q} of the AHM for various angles $\theta$ as shown in~\cref{fig:ovlap-exact-gs-and-approx-energy-mom-dist}(d)-(g).
We observe that the ground state has a large overlap with the lower branch of the two-body bound state of the corresponding statistical angle $\theta$.
As a consequence, due to the finite slope of the bound-state dispersion at this COM quasi-momentum, we can argue that the ground state $\ket{\Phi}$ prepares a moving 
bound state for $U=0$ to a good approximation.
The reason for the large overlap of the ground state $\ket{\Phi}$ with the two-body bound states at finite COM quasi-momentum can be attributed to two main factors:
(i) The exact ground state is predominantly localized on configurations with relatively small distances $m\leq 2$ between the two particles, and (ii) the fixed phase relationship of the ground state between these configurations, which depends on the statistical angle $\theta$, leads to a COM quasi-momentum distribution whose center is shifted to a finite value $q\approx -\theta$ around which the two-body bound states are also found.

Note that this contrasts with previous experiments, where initial states were given by
two-particle Fock states $\ket{\ldots 0110 \ldots}$ with sharp site occupation~\cite{2024_Kwan}.  These two-particle Fock states are completely delocalized in COM quasi-momentum space and
have equally populated the upper and lower branches of the two-body bound states with opposite slopes, such that a moving bound state could only be prepared by finite on-site interactions~\cite{2024_Kwan}. 
\subsubsection{COM position, width, and mean velocity}\label{subsubsec:COM-pos-vel}
To gain an understanding of the effect of the statistical angle $\theta$ on the asymmetric drift~[\cref{fig:com_motion_theta_t}], we will compute the time evolution of the initial state of the experimental protocol~$\ket{\Phi}$~\eqref{eq:AHM-ground-state-chiral-basis} perturbatively in powers of the evolution time. 
Then, to understand how directed transport builds up, we will compute the COM position,
the width and the mean velocity of this time-evolved state.

To this end, we first specify the relations for the COM position and the width of the time-evolved state and derive the COM velocity operator, which depends on the current operator in the AHM.
Due to the density-dependent tunneling matrix elements, the current operator in AHM 
will be affected by statistical angle $\theta$.
Generally, current operators for a finite lattice are defined via the continuity equation.
Thus, by studying the time evolution of the density operator $n_{j}$, we can identify a current operator.
To this end, we start with the Heisenberg equations of motion for the density operator,
\begin{align}
    \del_t \hat{n}_{j} &= -\frac{i}{\hbar} [\hat{n}_{j},\hat{H}]\nonumber \\
    &= \frac{iJ}{\hbar} \left(\hat{b}^{\dagger}_{j} e^{-i \theta \hat{n}_{j}} \hat{b}_{j-1}
    - \hat{b}^{\dagger}_{j+1} e^{-i \theta \hat{n}_{j+1}} \hat{b}_{j}  - \text{h.c.} \right),
    \label{eq:Heisenberg-eom}
\end{align}
where $\hat{H}$ is the AHM Hamiltonian \eqref{eq:AHM}.
In \cref{eq:Heisenberg-eom} we can now identify the current operator $\hat{j}_j$ for the AHM as follows
\begin{align}
    \del_t &\hat{n}_{j}=\hat{j}_j-\hat{j}_{j+1}, \nonumber \\
    \text{with} \ &\hat{j}_j=\frac{iJ}{\hbar} \left(\hat{b}^{\dagger}_{j} e^{-i \theta \hat{n}_{j}} \hat{b}_{j-1}
    - \text{h.c.} \right).
    \label{eq:current-op}
\end{align}    
The COM position operator $\hat{X}$ is given by
\begin{align}
    \hat{X}=\frac{1}{N}\sum_j j\,\hat{n}_j,
    \label{eq:COM-op}
\end{align}
by which we can then specify the corresponding COM velocity operator $\hat{V}=\del_t \hat{X}$, which reads
\begin{align}
    \hat{V} = \frac{1}{N}\sum_j j\, \del_t \hat{n}_j = \frac{1}{N}\sum_j j \,( \hat{j}_j- \hat{j}_{j+1}),
    \label{eq:COM-vel-op}
\end{align}
where $N$ denotes the total number of particles in the system.
The mean COM position (velocity) of a state in this system is then given 
by the expectation value of the COM position (velocity) operator $\ep{\hat{X}}$ ($\ep{\hat{V}}$).

We will also quantify the expansion by the width of the 
real space density profile, denoted as $(\Delta X)^2$, which is given by \cite{COM_pos_op} % Footnote added to Biblio.bib file
\begin{align}
    (\Delta X)^2 &= \frac{1}{N}\sum_j (j)^2\, \ep{\hat{n}_j} 
    -\bigg(\frac{1}{N}\sum_j j\, \ep{\hat{n}_j} \bigg)^2 \nonumber \\
    &= \frac{1}{N}\sum_j (j)^2\, \ep{\hat{n}_j} - \ep{\hat{X}}^2.
    \label{eq:COM-var-op}
\end{align}

\subsubsection{Time-dependent corrections}\label{subsec:time-dependent-corrections}
\begin{figure}[bt!]
    \centering
    \includegraphics[width=\columnwidth]{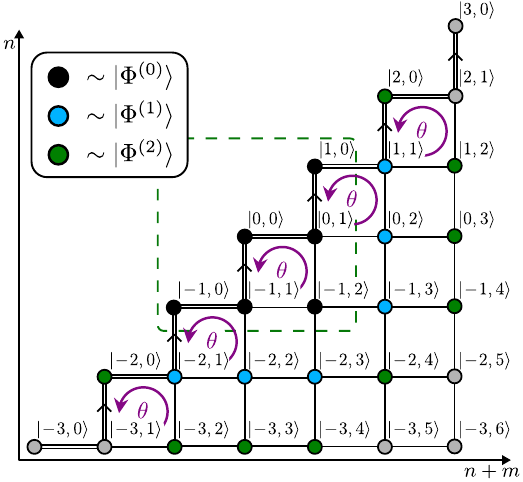}
    \caption{\textbf{Time-evolved perturbative corrections of the initial state 
    in the configuration space.}
    Perturbative time-dependent corrections to the initial state $\ket{\Phi}$~\eqref{eq:AHM-ground-state-chiral-basis} of the experiment in the configuration space (cf.~\cref{fig:relativecoord-AHM}).
    The black-filled circles (inside the green dashed square) 
    indicate the configurations that are present for the initial state configuration $\ket{\Phi}=\ket{\Phi^{(0)}}$.
    The light blue (green) filled circles indicate the configurations that are reached by
    first (second) order processes, i.e., which are present
    in the first (second) order correction 
    $\ket{\Phi^{(1)}}$ ($\ket{\Phi^{(2)}}$)~\eqref{eq:approx-state-1st-order}-\eqref{eq:approx-state-2nd-order}.
    The gray filled circles are not reached by the perturbative treatment 
    up to second order in $t$.}
    \label{fig:relativecoord-AHM-PT}
\end{figure}
We now compute the full time-dependent correction to second order in $t$ 
for the initial state $\ket{\Phi}$.
To this end, we apply the exact time evolution operator $\hat{U}(t)=\exp(-i\hat{H}t/\hbar)$ with $\hat{H}$ being the AHM Hamiltonian \eqref{eq:AHM}, 
onto the state $\ket{\Phi}$, 
\begin{align}
    \ket{\Phi(t)}={}&\hat{U}(t)\ket{\Phi} 
    =\left(\mathbbm{1}-i\frac{\hat{H}t}{\hbar}-\frac{(\hat{H}t)^2}{2\hbar^2}+\mathcal{O}(t^3)\right)\ket{\Phi}
    \nonumber \\
    ={}&\ket{\Phi^{(0)}}+\epsilon\ket{\Phi^{(1)}}+\epsilon^2\ket{\Phi^{(2)}}+\mathcal{O}(\epsilon^{3}),
    \label{eq:state-PT-expansion}
\end{align}
where $\epsilon=Jt/\hbar$ and where we have denoted the $k$-th order of correction in $t$ by a superscript $k$.
We note that the truncation of higher-order terms in \eqref{eq:state-PT-expansion}
generally leads to state corrections that are only normalized up to the order of the truncation.
However, for a truncated state at order $k=2$ in \eqref{eq:state-PT-expansion}, 
it can be shown that the state correction $\ket{\tilde{\Phi}(t)}\equiv \left(\mathbbm{1}-i\hat{H}t/\hbar-(\hat{H}t)^2/(2\hbar^2)\right)\ket{\Phi},$ is not only correctly normalized up to second but also up to third order in $t$, i.e., $\braket{\tilde{\Phi}(t)}{\tilde{\Phi}(t)}=1+\mathcal{O}(t^4)$, which suffices for our following analysis.
In the following, we will utilize the relative coordinate basis representation 
of the initial state $\ket{\Phi^{(0)}}=\sum_{n,m}d_{n,m}\ket{n,m}$~\eqref{eq:AHM-correct-gs-superpos-new}, 
with the analytically obtained coefficients $d_{n,m}$ 
from~\eqref{eq:AHM-ground-state-chiral-basis}, to compute the time-dependent corrections to the initial state~\eqref{eq:state-PT-expansion}. 
The higher-order time-dependent corrections to the initial state will necessarily lead to the population of configurations $\ket{n,m}$, 
which are not included in the initial state~\eqref{eq:AHM-correct-gs-superpos-new}.
The configurations, which are reached by the first and second order perturbative treatment, 
are illustrated in~\cref{fig:relativecoord-AHM-PT}.

We compute the first-order correction to the initial state and collect the expansion coefficients of the populated configurations, yielding 
\begin{widetext}
    \begin{align}
            \hat{U}(t)\ket{\Phi} ={}&\ket{\Phi^{(0)}}+ i\frac{Jt}{\hbar} 
            \bigg[ 
                d^{(0)}_{-1,2}\ket{-2,3}+d^{(0)}_{-1,2}\ket{-1,3}
                +d^{(0)}_{-1,1}\ket{-2,2}+(d^{(0)}_{0,1}+d^{(0)}_{-1,1})\ket{-1,2}+d^{(0)}_{0,1}\ket{0,2} \nonumber \\
                +&\sqrt{2}e^{i\theta}d^{(0)}_{-1,0}\ket{-2,1}
                +\left(\sqrt{2}d^{(0)}_{-1,0}+\sqrt{2}e^{i\theta}d^{(0)}_{0,0}
                +d^{(0)}_{-1,2}\right)\ket{-1,1}
                +\left(\sqrt{2}d^{(0)}_{0,0}+\sqrt{2}e^{i\theta}d^{(0)}_{1,0}
                +d^{(0)}_{-1,2}
                \right)\ket{0,1} \nonumber \\
                +&\sqrt{2}d^{(0)}_{1,0}\ket{1,1}
                +\sqrt{2}d^{(0)}_{-1,1}\ket{-1,0}
                +\left(\sqrt{2}d^{(0)}_{0,1}+\sqrt{2}e^{-i\theta}d^{(0)}_{-1,1}\right)\ket{0,0}
                +\sqrt{2}e^{-i\theta}d^{(0)}_{0,1}\ket{1,0}
                \bigg]
                +\mathcal{O}\left(\frac{Jt}{\hbar}\right)^2 \nonumber \\
               ={}& \ket{\Phi^{(0)}}+ \epsilon\ket{\Phi^{(1)}} +\mathcal{O}\left(\epsilon^2\right),
            \label{eq:approx-state-1st-order}
        \end{align}
    \end{widetext}
where we have added a superscript to the coefficients, i.e., $d^{(k)}_{n,m}$, indicating
the order $k$ from which perturbative correction they originate from. 
Here, this means $d^{(0)}_{n,m}\equiv d_{n,m}$ for the zero-th order coefficients of the initial state $\ket{\Phi^{(0)}}\equiv \ket{\Phi}$~\eqref{eq:AHM-ground-state-chiral-basis}, and for the first order we have $\ket{\Phi^{(1)}}=\sum_{n,m}d^{(1)}_{n,m}\ket{n,m}$, where the coefficients $d^{(1)}_{n,m}$ are, for instance, given by $d^{(1)}_{0,0}=i\sqrt{2}\left(d^{(0)}_{0,1}+e^{-i\theta}d^{(0)}_{-1,1}\right)$~\eqref{eq:approx-state-1st-order}.

The second-order correction to $\ket{\Phi}$ reads
\begin{widetext}
    \begin{align}
        \begin{split}
            \hat{U}(t)\ket{\Phi} ={}&\ket{\Phi^{(0)}}+ \epsilon\ket{\Phi^{(1)}}+\frac{i}{2}\left(\frac{Jt}{\hbar}\right)^2
            \bigg[ 
                d^{(1)}_{-2,3} \ket{-3,4}+ \left(d^{(1)}_{-2,3}+d^{(1)}_{-1,3}\right)\ket{-2,4}
                +d^{(1)}_{-1,3}\ket{-1,4}\\
                &{}+
                d^{(1)}_{-2,2} \ket{-3,3}
                +\left(d^{(1)}_{-2,2}+d^{(1)}_{-1,2}\right)\ket{-2,3}
                +\left(d^{(1)}_{0,2}+d^{(1)}_{-1,2}\right)\ket{-1,3}
                +d^{(1)}_{0,2}\ket{0,3}\\
                &{}+d^{(1)}_{-2,1}\ket{-3,2}+(d^{(1)}_{-2,1}+d^{(1)}_{-1,1}+d^{(1)}_{-2,3})\ket{-2,2}+\left(d^{(1)}_{-1,1}+d^{(1)}_{0,1}+d^{(1)}_{-2,3}+d^{(1)}_{-1,3}\right)\ket{-1,2}\\
                &{}+\left(d^{(1)}_{0,1}+d^{(1)}_{1,1}
                +d^{(1)}_{-1,3}\right)\ket{0,2}+d^{(1)}_{1,1}\ket{1,2}
                +\left(d^{(1)}_{-2,2}+\sqrt{2}e^{i\theta}d^{(1)}_{-1,0}\right)\ket{-2,1}\\
                &{}+\left(\sqrt{2}d^{(1)}_{-1,0}+\sqrt{2}e^{i\theta}d^{(1)}_{0,0}+d^{(1)}_{-1,2}+d^{(1)}_{-2,2}\right)\ket{-1,1}\\
                &{}+\left(\sqrt{2}d^{(1)}_{0,0}+\sqrt{2}e^{i\theta}d^{(1)}_{1,0}+d^{(1)}_{0,2}+d^{(1)}_{-1,2}\right)\ket{0,1}
                +\left(d^{(1)}_{0,2}+\sqrt{2}d^{(1)}_{1,0}\right)\ket{1,1}\\
                &{}+\sqrt{2}d^{(1)}_{-2,1}\ket{-2,0}+\left(\sqrt{2}e^{-i\theta}d^{(1)}_{-2,1}+\sqrt{2}d^{(1)}_{-1,1}\right)\ket{-1,0}
                +\left(\sqrt{2}e^{-i\theta}d^{(1)}_{-1,1}+\sqrt{2}d^{(1)}_{0,1}\right)\ket{0,0}\\
                &{}+\left(\sqrt{2}e^{-i\theta}d^{(1)}_{0,1}+\sqrt{2}d^{(1)}_{1,1}\right)\ket{1,0}+\sqrt{2}e^{-i\theta}d^{(1)}_{1,1}\ket{2,0} 
                \bigg]
                +\mathcal{O}\left(\frac{Jt}{\hbar}\right)^3\\
                ={}&\ket{\Phi^{(0)}}+ \epsilon\ket{\Phi^{(1)}} + \epsilon^2\ket{\Phi^{(2)}} 
                +\mathcal{O}\left(\epsilon^3\right),
            \end{split}
            \label{eq:approx-state-2nd-order}
        \end{align}
    \end{widetext}
where now for the second order we have $\ket{\Phi^{(2)}}=\sum_{n,m}d^{(2)}_{n,m}\ket{n,m}$, where the coefficients $d^{(2)}_{n,m}$ are, for instance, given by $d^{(2)}_{0,0}=i\frac{\sqrt{2}}{2}\left(e^{-i\theta}d^{(1)}_{-1,1}+d^{(1)}_{0,1}\right)$~\eqref{eq:approx-state-2nd-order}.

Equipped with these perturbative expressions \eqref{eq:approx-state-1st-order}-\eqref{eq:approx-state-2nd-order}, we will now first compute the time-dependent corrections to the COM
position $X=\ep{\hat{X}}$~\eqref{eq:COM-op}, i.e.,
    \begin{align}
        X = X^{(0)}+\epsilon X^{(1)}+\epsilon^2 X^{(2)}+\mathcal{O}(\epsilon^3).
        \label{eq:COM-pos-PT}
    \end{align}
The individual orders $X^{(n)}$ in~\eqref{eq:COM-pos-PT} are given as follows.
The zeroth order $X^{(0)}$ reads 
    \begin{align}
            X^{(0)}=\frac{1}{N}\sum_j j\,\matel{\Phi^{(0)}}{\hat{n}_j}{\Phi^{(0)}},
        \label{eq:COM-pos-PT-order-0-th}
    \end{align}
and the first order $X^{(1)}$ reads 
    \begin{align}
        X^{(1)}=\frac{1}{N}\sum_j j\left( \matel{\Phi^{(0)}}{\hat{n}_j}{\Phi^{(1)}}+\text{c.c.} \right),
            \label{eq:COM-pos-PT-order-1st}
    \end{align}
where c.c.~stands for complex conjugate.
Finally, the second order $X^{(2)}$ reads
    \begin{align}
        X^{(2)}&=\frac{1}{N}\sum_j j\,\Big(\matel{\Phi^{(1)}}{\hat{n}_j}{\Phi^{(1)}} \nonumber \\
            &
            +\big[\matel{\Phi^{(2)}}{\hat{n}_j}{\Phi^{(0)}}+\text{c.c.} \big]\Big).
        \label{eq:COM-pos-PT-order-2nd}
    \end{align}
2nd order.
In \eqref{eq:COM-pos-PT-order-0-th}-\eqref{eq:COM-pos-PT-order-2nd}, 
the lattice site index $j$ runs from $-3$ to $3$, since only Fock states of 
the form $\ket{n_{-3},n_{-2},n_{-1},n_0,n_1,n_2,n_3}$ are relevant in the 2nd order perturbative treatment considered here~[\cref{fig:relativecoord-AHM-PT}].

By explicitly using the definition of the coefficients $d^{(0)}_{n,m}=d_{n,m}$~\eqref{eq:AHM-ground-state-chiral-basis}, we arrive at the following contributions for the COM position. 
In the zeroth order \eqref{eq:COM-pos-PT-order-0-th} we find 
    \begin{align}
        X^{(0)} = \frac{1}{2}\Bigg[-&\left(2\abs{d^{(0)}_{-1,0}}^2
        +\abs{d^{(0)}_{-1,1}}^2 
        +\abs{d^{(0)}_{-1,2}}^2 \right)
        \nonumber \\
        +&\left(2\abs{d^{(0)}_{1,0}}^2+\abs{d^{(0)}_{0,1}}^2 
        +\abs{d^{(0)}_{-1,2}}^2
        \right)\Bigg]=0.
    \label{eq:COM-pos-PT-0th-order}
     \end{align}

\begin{figure}[t!]
    \centering
    \hspace*{-.5cm}  % Adjust this value as needed to shift the figure to the left
    \includegraphics[width=1\columnwidth]{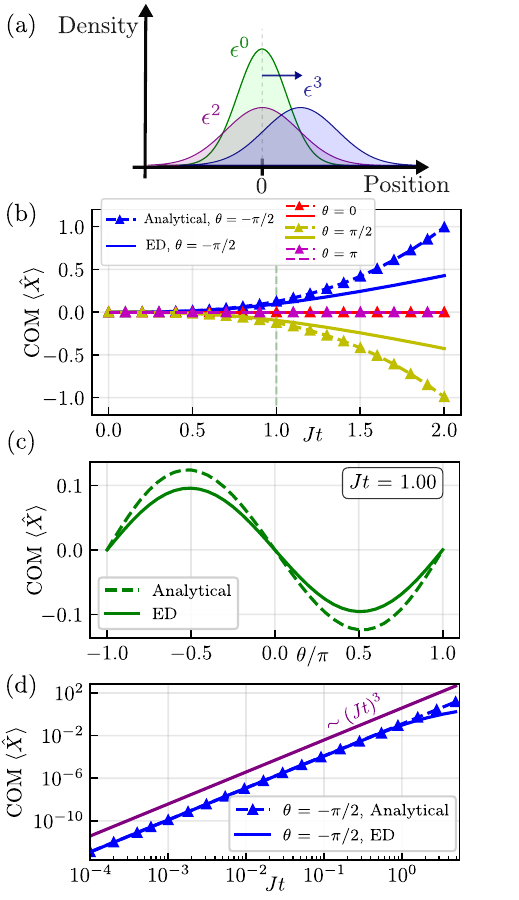}
    \caption{\textbf{Analytical COM motion from time-dependent perturbation theory.}
        (a) Illustration of density profile as a function of time.
        At zeroth order ($\epsilon^0$), the density profile is static and symmetric around the center of the lattice with a finite width.
        At second order ($\epsilon^2$), the density profile broadens symmetrically while its COM position remains unchanged.
        At third order ($\epsilon^3$), the density profile starts to move with a finite COM velocity.
        (b) COM position $X(t)$ obtained from third-order time-dependent perturbation theory~\eqref{eq:COM-PT-Final} (triangles) and from exact diagonalization (ED) (solid lines)
        (cf.~\cref{fig:com_motion_theta_t}(a)) as a function of time
        for various statistical angles $\theta \in \{-\pi/2,\,0,\,\pi/2,\,\pi\}$ (colors).
        (c) COM position $X(t)$ obtained from third-order time-dependent perturbation theory~\eqref{eq:COM-PT-Final} (dashed) and from ED (solid) 
        (cf.~\cref{fig:com_motion_theta_t}(b)) as a function of statistical angle $\theta$ at a fixed evolution time of $Jt=1$.
        (d) COM position $X(t)$ from third-order time-dependent perturbation theory~\eqref{eq:COM-PT-Final} (triangles) and from ED (solid) for $\theta=-\pi/2$ in a semilogarithmic plot for short evolution times.
        The purple solid line indicates a cubic power law $\sim (Jt)^3$.
             }
    \label{fig:COM_PT_theta_t_log}
\end{figure}

A lengthy but straightforward calculation also gives us the first- and second-order correction,
$X^{(1)}$~\eqref{eq:COM-pos-PT-order-1st} and $X^{(2)}$~\eqref{eq:COM-pos-PT-order-2nd} respectively, which read
\begin{widetext}
    \begin{align}
        X^{(1)} = \frac{1}{2}\Bigg[-&
        \left(i \left\{2\sqrt{2}\,d^{(0)}_{-1,1}(d^{(0)}_{-1,0})^{*}
        +\left(\sqrt{2}d^{(0)}_{-1,0}+\sqrt{2}e^{i\theta}d^{(0)}_{0,0}+d^{(0)}_{-1,2}\right)(d^{(0)}_{-1,1})^{*}\right\} +\text{c.c.} \right) \nonumber \\
        +&\left(i \left\{2\sqrt{2}e^{-i\theta}\,d^{(0)}_{0,1}(d^{(0)}_{1,0})^{*}
        +\left(\sqrt{2}d^{(0)}_{0,0}+\sqrt{2}e^{i\theta}d^{(0)}_{1,0}+d^{(0)}_{-1,2}\right)(d^{(0)}_{0,1})^{*}\right\} +\text{c.c.} \right)
        \Bigg]= 0,
        \label{eq:COM-pos-PT-1st-order}
    \end{align}
and 
\begin{align}
        X^{(2)} = {}&\frac{1}{2}\Bigg[-2\left(2\abs{d^{(0)}_{-1,0}}^2 +\abs{d^{(0)}_{-1,1}}^2+\abs{d^{(0)}_{-1,2}}^2 \right) \nonumber \\
        &-1\Bigg(\bigg(4\abs{d^{(0)}_{-1,1}}^2 +2\abs{d^{(0)}_{-1,0}}^2+
        \abs{\sqrt{2}(d^{(0)}_{-1,0}+e^{i\theta}d^{(0)}_{0,0})+d^{(0)}_{-1,2}}^2
        +\abs{d^{(0)}_{0,1}+d^{(0)}_{-1,1}}^2+\abs{d^{(0)}_{-1,2}}^2\bigg) \nonumber \\
        &+\bigg[\frac{-i}{2}\bigg\{\sqrt{2}\left(e^{i\theta}(d^{(1)}_{-2,1})^{*}+(d^{(1)}_{-1,1})^{*}\right)2 d^{(0)}_{-1,0}
         +\left(\sqrt{2}\left((d^{(1)}_{-1,0})^{*}+e^{-i\theta}(d^{(1)}_{0,0})^{*}\right)+(d^{(1)}_{-1,2})^{*}+(d^{(1)}_{-2,2})^{*}\right) d^{(0)}_{-1,1} \nonumber \\
         &-\left((d^{(1)}_{-1,1})^{*}+(d^{(1)}_{0,1})^{*}+(d^{(1)}_{-2,3})^{*}+(d^{(1)}_{-1,3})^{*}\right)d^{(0)}_{-1,2}
         \bigg\}+\text{c.c.} \bigg]
        \Bigg)\nonumber \\
        &+1\Bigg(\left(4\abs{d^{(0)}_{0,1}}^2 +2\abs{d^{(0)}_{1,0}}^2+2\abs{d^{(0)}_{0,0}+e^{i\theta}d^{(0)}_{1,0}}^2
        +\abs{d^{(0)}_{0,1}+d^{(0)}_{-1,1}}^2\right) \nonumber \\
        &{}+\bigg[\frac{-i}{2}\bigg\{\sqrt{2}\left(e^{i\theta}(d^{(1)}_{0,1})^{*}+(d^{(1)}_{1,1})^{*}\right)2 d^{(0)}_{1,0}
        +\left(\sqrt{2}\left((d^{(1)}_{0,0})^{*}+e^{-i\theta}(d^{(1)}_{1,0})^{*}\right)+(d^{(1)}_{0,2})^{*}+(d^{(1)}_{-1,2})^{*}\right) d^{(0)}_{0,1} \nonumber \\
         &-\left((d^{(1)}_{-1,1})^{*}+(d^{(1)}_{0,1})^{*}+(d^{(1)}_{-2,3})^{*}+(d^{(1)}_{-1,3})^{*}\right)d^{(0)}_{-1,2}
        \bigg\}+\text{c.c.} \bigg]
        \Bigg)\nonumber \\
        &+2\left(2\abs{d^{(0)}_{1,0}}^2 +\abs{d^{(0)}_{0,1}}^2 +\abs{d^{(0)}_{-1,2}}^2\right)
        \Bigg] =0.
    \label{eq:COM-pos-PT-2nd-order}
    \end{align}
We observe that the COM position $X$ is not affected by the perturbative treatment 
up to second order in $t$, i.e., $X^{(0)}=X^{(1)}=X^{(2)}=0$, implying 
that the initial state $\ket{\Phi}$ will only deviate from its initial position 
at third or higher order in $t$.

Next, we compute time-dependent corrections to the width of the time-evolved state $(\Delta X)^2$ \eqref{eq:COM-var-op},
i.e.,
    \begin{align}
            (\Delta X)^2 
            = &((\Delta X)^2)^{(0)}+\epsilon ((\Delta X)^2)^{(1)}+\epsilon^2 ((\Delta X)^2)^{(2)}\nonumber \\
            &+\mathcal{O}(\epsilon^3).
        \label{eq:COM-var-op-PT}
    \end{align}
The zeroth order is given by
    \begin{align}
        ((\Delta X)^2)^{(0)} = {}&\frac{1}{N}\sum_j j^2\,\matel{\Phi^{(0)}}{\hat{n}_j}{\Phi^{(0)}}
        -\bigg(\frac{1}{N}\sum_j j\, \matel{\Phi^{(0)}}{\hat{n}_j}{\Phi^{(0)}} \bigg)^2 \nonumber \\
        ={}&\frac{1}{N}\sum_j j^2\,\matel{\Phi^{(0)}}{\hat{n}_j}{\Phi^{(0)}}
        -\left(X^{(0)}\right)^2
        \nonumber \\
        ={}& \frac{1}{2}\Bigg[(-1)^2\left(2\abs{d^{(0)}_{-1,0}}^2
        +\abs{d^{(0)}_{-1,1}}^2 
        \abs{d^{(0)}_{-1,2}}^2 \right)
        (+1)^2\left(2\abs{d^{(0)}_{1,0}}^2+\abs{d^{(0)}_{0,1}}^2 
        \abs{d^{(0)}_{-1,2}}^2
        \right)\Bigg]-0 \nonumber \\
        ={}&\frac{\abs{c}^{2}+11\abs{c}+2\mathrm{Re}[c]}{4\abs{c}(\abs{c}+5)} =
        \frac{7+8\cos\theta+11\sqrt{5+4\cos\theta}}{4\sqrt{5+4\cos\theta}(5+\sqrt{5+4\cos\theta})}.
        \label{eq:COM-var-PT-0th-order}
    \end{align}
At first order, the width is given by,
\begin{align}
        ((\Delta X)^2)^{(1)} &= \frac{1}{N}\sum_j j^2\big( \matel{\Phi^{(0)}}{\hat{n}_j}{\Phi^{(1)}}+\text{c.c.} \big)\nonumber \\
        &= \frac{1}{2}\Bigg[(-1)^2\left(i\left\{2\sqrt{2}\,d^{(0)}_{-1,1}(d^{(0)}_{-1,0})^{*}
        +\left(\sqrt{2}d^{(0)}_{-1,0}+\sqrt{2}e^{i\theta}d^{(0)}_{0,0}+d^{(0)}_{-1,2}\right)(d^{(0)}_{-1,1})^{*}\right\} +\text{c.c.} \right) 
        \nonumber \\
        &\quad\quad \quad \! (+1)^2
        \left(i\left\{2\sqrt{2}e^{-i\theta}\,d^{(0)}_{0,1}(d^{(0)}_{1,0})^{*}
        +\left(\sqrt{2}d^{(0)}_{0,0}+\sqrt{2}e^{i\theta}d^{(0)}_{1,0}+d^{(0)}_{-1,2}\right)(d^{(0)}_{0,1})^{*}\right\} +\text{c.c.} \right)
        \Bigg]
        = 0 .
        \label{eq:COM-var-PT-1st-order}
    \end{align}
Eventually, for the width of the wave packet in second order, we find
    \begin{align}
        ((\Delta X)^2)^{(2)} = {}&\frac{1}{N}\sum_j j^2\,\Big(\matel{\Phi^{(1)}}{\hat{n}_j}{\Phi^{(1)}} 
        +\big[\matel{\Phi^{(0)}}{\hat{n}_j}{\Phi^{(2)}}+\text{c.c.} \big]\Big) 
        -\bigg(\frac{1}{N}\sum_j j\,\big(\matel{\Phi^{(0)}}{\hat{n}_j}{\Phi^{(1)}}+\text{c.c.} \big) \bigg)^2 \nonumber  \\
        = {}&\frac{1}{N}\sum_j j^2\,\Big(\matel{\Phi^{(1)}}{\hat{n}_j}{\Phi^{(1)}} 
        +\big[\matel{\Phi^{(0)}}{\hat{n}_j}{\Phi^{(2)}}+\text{c.c.} \big]\Big) 
        -\left(X^{(1)}\right)^2 \nonumber  \\
         ={}&\frac{3(2\abs{c}^{2}+11\abs{c}-3)}{4\abs{c}(\abs{c}+5)} 
        = \frac{21 + 24\cos\theta + 33\sqrt{5+4\cos\theta}}
            {20+16\cos\theta + 20\sqrt{5+4\cos\theta}}.
        \label{eq:COM-var-PT-2nd-order}
    \end{align}
\end{widetext}
The finite width of the density profile of the initial state is reflected in the 
non-zero value of the width $(\Delta X)^2$ in zeroth order \eqref{eq:COM-var-PT-0th-order}.
We find that the width $(\Delta X)^2$ only starts to deviate from its initial value at second order
in $t$ \eqref{eq:COM-var-PT-2nd-order}.
Thus, we can conclude that the cloud of particles in the experiment first broadens symmetrically
at second order in $t$ while its COM position $X$ remains stationary up to this order,
related to $((\Delta X)^2)^{(2)}\neq0$ and $X^{(0)}=X^{(1)}=X^{(2)}=0$ respectively,
and only then actually starts to move at cubic (or higher) order in $t$, as illustrated in~\cref{fig:COM_PT_theta_t_log}(a).

At last, we compute the time-dependent corrections to the 
mean velocity $V=\ep{\hat{V}}$ \eqref{eq:COM-vel-op},
    \begin{align}
        \begin{split}
            V = V^{(0)}+\epsilon V^{(1)}+\epsilon^2 V^{(2)}+\mathcal{O}(\epsilon^3).
        \end{split}
        \label{eq:COM-vel-PT}
    \end{align}
Here, we note that the relationship between the COM velocity and position, $\del_t X=V$~\eqref{eq:COM-vel-op}, also holds at each order of $t$ in our perturbative expansion~(\eqref{eq:COM-pos-PT},\eqref{eq:COM-vel-PT}), i.e.,
    \begin{align}
        \frac{J}{\hbar}(n+1) X^{(n+1)}=V^{(n)}, \ \forall n\in \mathbb{N}.
        \label{eq:COM-vel-pos-rel-order}
\end{align}
Thus, since the COM position in first~\eqref{eq:COM-pos-PT-1st-order} and second order~\eqref{eq:COM-pos-PT-2nd-order} vanishes, we can directly conclude that the mean velocity $V$ also vanishes in zeroth and first order, respectively, i.e.,
\begin{align}
            V^{(0)} &=\frac{J}{\hbar}X^{(1)}= \frac{1}{N}\sum_j   j \,\matel{\Phi^{(0)}}{(\hat{j}_j- \hat{j}_{j+1})}{\Phi^{(0)}}=0.
        \label{eq:COM-vel-PT-0th-order}
    \end{align}
and
    \begin{align}
            V^{(1)} &=\frac{J}{\hbar}2 X^{(2)} \nonumber \\
            &= \frac{1}{N}\sum_j j\,\big( \matel{\Phi^{(0)}}{(\hat{j}_j- \hat{j}_{j+1})}{\Phi^{(1)}}+\text{c.c.} \big) =0.
        \label{eq:COM-vel-PT-1st-order}
    \end{align}
Now, the second-order contribution to the mean velocity $V$ has to be computed explicitly (since we cannot directly infer it from $X^{(3)}$ as we did not compute the perturbative state corrections beyond second order). 
Eventually, after some tedious but straightforward calculations, we arrive at 
    \begin{align}
            V^{(2)}  &=\frac{1}{N}\sum_j j\,\Big(\matel{\Phi^{(1)}}{(\hat{j}_j- \hat{j}_{j+1})}{\Phi^{(1)}} \nonumber \\
            &+\big[\matel{\Phi^{(0)}}{(\hat{j}_j- \hat{j}_{j+1})}{\Phi^{(2)}}+\text{c.c.} \big]\Big) \nonumber \\
            &=
            -\frac{J}{\hbar}\sin{\theta} 
            \Bigg[
            \frac{1}{\sqrt{\abs{c}+5}} 
            \Bigg]\nonumber \\
            &=
            -\frac{J}{\hbar}
            \frac{\sin{\theta}}{\,\bigl(4\cos\theta+10(3+\sqrt{5+4\cos\theta})\bigr)^{1/4}}
        \label{eq:COM-vel-PT-2nd-order}
    \end{align}
At quadratic order in $t$, we find a non-zero mean velocity $V$.
From the relationship between the mean velocity and position~\eqref{eq:COM-vel-pos-rel-order}, 
we can directly deduce the cubic contribution to the COM position $X^{(3)}$ from the quadratic contribution to the mean velocity $V^{(2)}$~\eqref{eq:COM-vel-PT-2nd-order}, 
resulting in
    \begin{align}
            X^{(3)} &=
            \frac{1}{3}
            \Bigg[
                -\sin{\theta} 
                \Bigg(\frac{1}{\sqrt{\abs{c}+5}} \Bigg)
                \Bigg]
            \nonumber \\
            &=
            \frac{-\sin{\theta}}{3\bigl(4\cos\theta+10(3+\sqrt{5+4\cos\theta})\bigr)^{1/4}}.
        \label{eq:COM-pos-PT-3rd-order}
    \end{align}

Altogether, for our initial state $\ket{\Phi}$, 
we find the following time-dependent perturbative corrections to the 
COM position $X$~\eqref{eq:COM-pos-PT}, the width $(\Delta X)^2$~\eqref{eq:COM-var-op-PT}, and the mean velocity $V$~\eqref{eq:COM-vel-PT} respectively,
\begin{align}
    X &=\left(
        \frac{-\sin{\theta}}{3\sqrt{\abs{c}+5}}
        \right)
        \epsilon^3+\mathcal{O}(\epsilon^4),
        \label{eq:COM-PT-Final}\\
        (\Delta X)^2 &=\left(\frac{\abs{c}^{2}+11\abs{c}+2\mathrm{Re}[c]}{4\abs{c}(\abs{c}+5)} \right)
        \nonumber \\
        &+\left(\frac{3(2\abs{c}^{2}+11\abs{c}-3)}{4\abs{c}(\abs{c}+5)}\right)\epsilon^2
        +\mathcal{O}(\epsilon^3),
        \label{eq:Var-PT-Final}\\
        V &=\left(-\frac{J}{\hbar}
        \frac{\sin{\theta}}{\sqrt{\abs{c}+5}} \right)\epsilon^2 +\mathcal{O}(\epsilon^3),
        \label{eq:Vel-PT-Final}
    \end{align}
where $\epsilon=Jt/\hbar$, $c=1+2e^{i\theta}$, and $\abs{c}=\sqrt{5+4\cos(\theta)}$~\eqref{eq:AHM-eigenvectors-AAdagger} again.
In~\cref{fig:COM_PT_theta_t_log}(b) and (c), together with the numerically exact results obtained from ED, we depict the time-dependent COM position $X(t)$~\eqref{eq:COM-PT-Final} from third-order perturbation theory as a function of time for various statistical angles $\theta$ and as a function of $\theta$ at a fixed evolution time, respectively.
We find excellent agreement with the ED results for moderately short evolution times $Jt\lesssim  1$, as expected from a time-dependent perturbative treatment.
Moreover, in~\cref{fig:COM_PT_theta_t_log}(d), we also show for $\theta=-\pi/2$ the perturbative COM position $X(t)$~\eqref{eq:COM-PT-Final} together with the numerically exact ED result in a semilogarithmic plot. We observe the same slope for the exact and perturbative COM position $X(t)$, implying that both follow the same cubic power-law behavior for short evolution times.

The fact that the cloud of the particles first becomes broader in second order~\eqref{eq:Var-PT-Final} before acquiring a finite COM drift in third order implies that a finite COM velocity emerges in the same order as the broadening~\eqref{eq:Vel-PT-Final}, i.e., that the two are directly related~[\cref{fig:COM_PT_theta_t_log}(a)].
This can be understood intuitively from a simple mean-field argument. 
Replacing the number-dependent tunneling phase $\theta \hat{n}_{j+1}$ 
in the Hamiltonian~\eqref{eq:AHM} by a mean-field phase $\theta\rho$, with mean occupation per site, $\rho$, causes a shift of the dispersion relation by $-\theta \rho$. 
For the initial state, the wave function is centered around the minimum of this shifted mean-field dispersion relation. 
Now, when the cloud expands, $\rho$ decreases, so that the dispersion relation shifts, 
leaving the cloud displaced with respect to the minimum of the dispersion relation and, 
thus, acquiring a finite group velocity. 
In other words, the expansion of the cloud leads to a time-dependent change in the effective gauge field (vector potential) $A\sim \rho\theta$, which then induces an electric force on the particles, resulting in the aforementioned finite group velocity.
A very similar observation was made in a recent experiment in Barcelona~\cite{2022_Frolian}, where a density-dependent electric force emerged \cite{2022_Frolian,2022_Chisholm}.
This connection is further corroborated by the deflection of the cloud at a barrier discussed in the main text, showing the chiral nature of the AHM cloud as in the Barcelona experiment~\cite{2022_Frolian}.

Beyond the mean-field level, a more precise picture of this reasoning in the context of two particles is that the relative phases of the initial wave function in the COM direction (which essentially determine where in quasi-momentum space the state is centered) are adapted to provide a state at rest, taking into account the finite Peierls phases present at short relative distances $m\leq 1$~(cf.~\cref{fig:relativecoord-AHM}). 
Now, when broadening in the relative direction $m$ after the walls are released, the state overlaps with a region without Peierls phases, where the phase profile no longer corresponds to a state at rest in the COM direction, thus leading to a finite group velocity.

\putbib % Inserts references for this unit
\end{bibunit}


\begin{thebibliography}{46}
\bibitem [{dou()}]{doublon}%
  \BibitemOpen
  \href@noop {} {\bibinfo {title} {{The doublon fraction is equivalently
  obtained by summing the diagonal elements $\Gamma_{ii}$ over all
  sites.}}}\BibitemShut {Stop}%
\end{thebibliography}
\end{document}